\documentclass[twoside,11pt,preprint]{article}

\usepackage[abbrvbib]{dmlr2e}
\usepackage{hyperref}       % hyperlinks
\usepackage{url}            % simple URL typesetting
\usepackage{booktabs}       % professional-quality tables
\usepackage{multirow}
\usepackage{amsmath}
\usepackage{xcolor}  
\usepackage{subcaption}
\usepackage{algorithm}
\usepackage[noend]{algpseudocode}
\usepackage{listings}

\definecolor{codebg}{rgb}{0.97,0.97,0.97}

\lstdefinestyle{mystyle}{
    backgroundcolor=\color{codebg},
    basicstyle=\ttfamily\footnotesize,
    breaklines=true,
    breakatwhitespace=true,
    keepspaces=true,
    numbers=left,
    numberstyle=\tiny\color{gray},
    numbersep=5pt,
    showspaces=false,
    showstringspaces=false,
    showtabs=false,
    tabsize=2,
    captionpos=b
}

\usepackage{lastpage}
\dmlrheading{23}{2026}{1-\pageref{LastPage}}{1/26; Revised XXX}{XXX}{21-0000}{Rebecca Salganik, et al.} % Insert the dates and author names. This is not relevant if it is yet being reviewed, i.e., \documentclass[twoside,11pt, preprint]{article}
\def\openreview{\url{https://music-sem-web.vercel.app/}} % Insert correct link to OpenReview for camera-ready version
\ShortHeadings{MusicSem}{Rebecca Salganik, et al. (2026)}
\firstpageno{1}

\begin{document}

\title{MusicSem: A Semantically Rich Language--Audio Dataset \\ of Natural Music Descriptions}

\author{\name Rebecca Salganik$^{1}$ \email rsalgani@ur.rochester.edu \\
        \name Teng Tu$^{2}$ \\
         \name Fei-Yueh Chen$^{1}$ \\
        \name Xiaohao Liu$^{2}$ \\
        \name Keifeng Lu$^{1}$ \\
        \name Ethan Luvisia$^{1}$ \\
     \name Zhiyao Duan$^{1}$ \\
     \name Guillaume Salha-Galvan$^{3}$ \\
     \name Anson Kahng$^{1}$ \\
     \name Yunshan Ma$^{4}$ \\
      \name Jian Kang$^{5}$ \\
       \addr $^{1}$University of Rochester, NY, USA \\
       \addr $^{2}$National University of Singapore, Singapore \\     
       \addr $^{3}$SJTU Paris Elite Institute of Technology, Shanghai, China \\
       \addr $^{4}$Singapore Management University, Singapore \\
       \addr $^{5}$Mohamed bin Zayed University of Artificial Intelligence, Abu Dhabi, UAE       
}

\editor{XXX}

\maketitle

\begin{abstract}
Music representation learning is central to music information retrieval and generation. While recent advances in multimodal learning have improved alignment between text and audio for tasks such as cross-modal music retrieval, text-to-music generation, and music-to-text generation, existing models often struggle to capture users’ expressed intent in natural language descriptions of music. This observation suggests that the datasets used to train and evaluate these models do not fully reflect the broader and more natural forms of human discourse through which music is described. In this paper, we introduce MusicSem, a dataset of 32,493 language–audio pairs derived from organic music-related discussions on the social media platform Reddit. Compared to existing datasets, MusicSem captures a broader spectrum of musical semantics, reflecting how listeners naturally describe music in nuanced and human-centered ways. To structure these expressions, we propose a taxonomy of five semantic categories: descriptive, atmospheric, situational, metadata-related, and contextual. In addition to the construction, analysis, and release of MusicSem, we use the dataset to evaluate a wide range of multimodal models for retrieval and generation, highlighting the importance of modeling fine-grained semantics. Overall, MusicSem serves as a novel semantics-aware resource to support future research on human-aligned multimodal music representation learning.
\end{abstract}


\section{Introduction}
\label{sec:introduction}

\paragraph{\textbf{Context and Motivation.}}
Music representation learning~\citep{Schedl_music_2014,muller_fundamentals_2015,hernandez2022music} underpins a wide range of downstream music-related tasks, including music categorization~\citep{oramas_multi_2017,mccallum_supervised_2022,yuan_marble_2023}, generation~\citep{hernandez2022music,gardner_llark_2024,liu_music_2024}, and recommendation~\citep{van2013deep,deldjoo2024content,salganik_larp_2024}. Early research in this area largely focused on audio-centric approaches, relying on handcrafted features or learned audio representations to model musical content~\citep{lin_exploiting_2011,oramas_multi_2017,bogdanov_mtg_2019}. More recently, advances in multimodal learning have enabled the joint modeling of audio and natural language descriptions of music, leading to substantial progress in tasks such as cross-modal music retrieval~\citep{wu_largescale_2023,girdhar_imagebine_2023,wu_clamp3_2025}, music-to-text generation~\citep{liu_music_2024,doh_lpmusiccaps_2023,wu_futga_2024}, and text-to-music generation~\citep{agostinelli_musiclm_2023,copet_simple_2023,evans_fast_2024,liu_audioldm2_2024}.

Effective language--audio multimodal music representation learning requires understanding how musical meaning and intent are expressed in natural language~\citep{ronchini_paguri_2024,zang_interpretation_2024}. In particular, capturing the nuances that contextualize a listening experience in textual descriptions is crucial. Consider, for example, a text-to-music generation or retrieval model and the following two descriptions used as input: \textit{``This song is a ballad. It contains guitar, male vocals, and a piano. It sounds like something I would listen to at church.''} versus \textit{``This song is a ballad. It contains guitar, male vocals, and a piano. It sounds like something I would listen to during a psychedelic experience.''} Although both descriptions specify identical musical attributes, the situational context leads to different expectations regarding the audio to be generated or retrieved.

However, recent research has shown that multimodal music representation learning models still often struggle to capture users' expressed intent in natural language descriptions of music~\citep{ronchini_paguri_2024,zang_interpretation_2024}. This observation points to limitations in existing language--audio datasets~\citep{agostinelli_musiclm_2023,manco_describer_2023}, which may not fully reflect the broader and more natural forms of human discourse through which music is described.
Furthermore, while professional musicians tend to rely on descriptive and technical language when engaging with music, non-expert listeners often express their experiences using more abstract and subjective semantic content~\citep{gromko_perceptual_1993,bainbridge_how_2003}. As a result, commonly used musician-annotated datasets for music representation learning may be overly curated and insufficiently representative of everyday listening experiences \citep{agostinelli_musiclm_2023}. Overall, this gap motivates the need for textual annotations that encompass a broader range of natural music descriptions.

\paragraph{\textbf{Contributions.}}
In this paper, we begin by formalizing the elements that contextualize a listening experience in textual descriptions, which we refer to as musical semantics~\citep{levy_learning_2008,nam_deep_2019,choi_prediction_2020}. We introduce a taxonomy that distinguishes five types of semantic captions: descriptive, contextual, situational, atmospheric, and metadata-based. We further show that many competitive multimodal music representation learning models for generation and retrieval, when trained on existing language--audio datasets, lack sensitivity to these semantic distinctions. %We also argue that this limitation tends to be exacerbated by the increasing use of large language model (LLM)-generated datasets, as well as by the absence of standardized evaluation protocols for generative and retrieval tasks.

\begin{figure}[t]
    \centering
    \includegraphics[width=\linewidth]{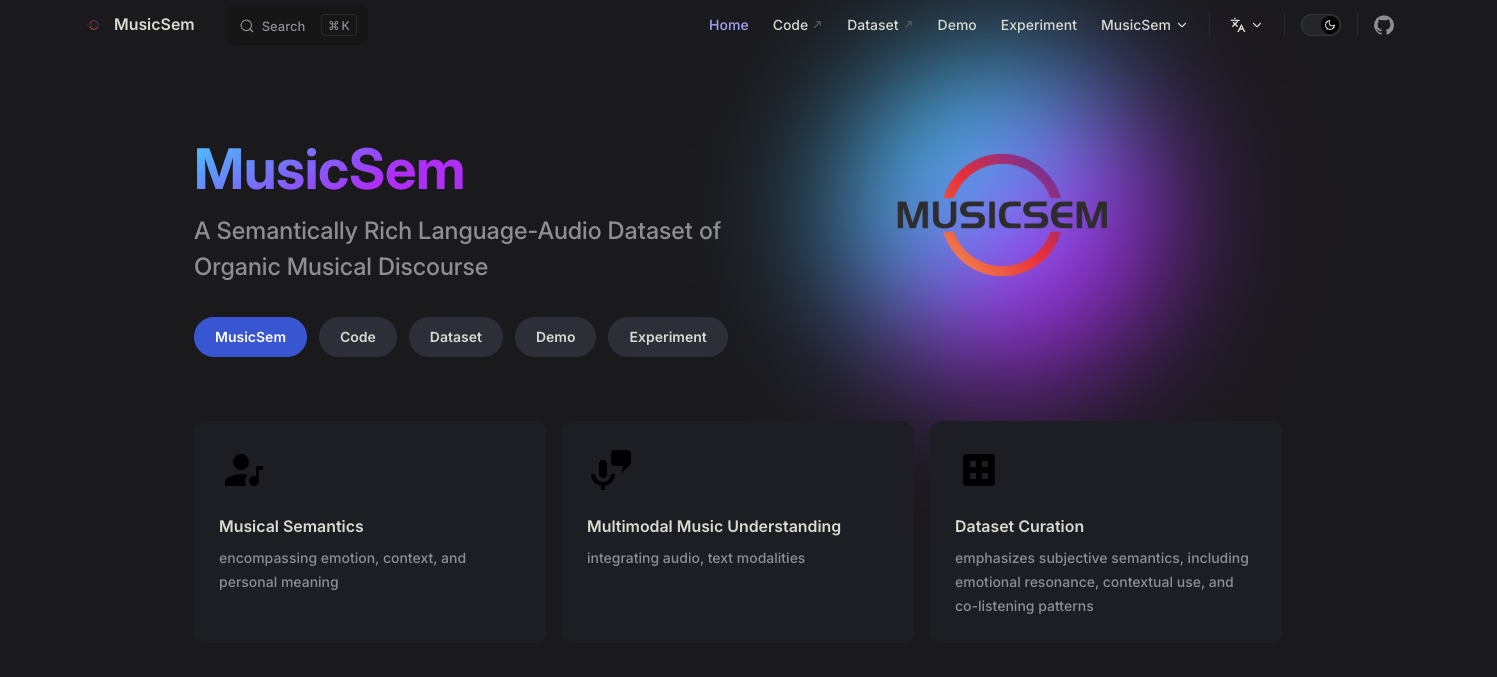}
\caption{The MusicSem website provides access to the full dataset, detailed documentation, visualizations, and source code for data construction and experiments at: \url{https://music-sem-web.vercel.app/}.}
    \label{fig:website}
\end{figure}

Motivated by these observations, we introduce MusicSem, a semantically rich language--audio dataset derived from organic music discussions on the social media platform Reddit. The dataset comprises 32,493 language--audio music description pairs, with textual annotations that express not only descriptive attributes of music, but also emotional resonance, contextual and situational usage, and co-listening patterns. MusicSem distinguishes itself by capturing a broader spectrum of musical semantics than prior datasets used for multimodal model training and evaluation.
We detail the entire construction pipeline of MusicSem, including our design choices and motivations, as well as the final dataset statistics and key characteristics.
Taking ethical and legal considerations seriously, we also present the key safeguards we adopted with respect to user privacy and anonymity, platform compliance, and music copyright.
As illustrated in Figure~\ref{fig:website}, the final MusicSem dataset is publicly available online, together with the complete source code for dataset construction.

Third, we perform an extensive evaluation of various cross-modal music retrieval, text-to-music generation, and music-to-text generation models using MusicSem. Our evaluation yields several key insights, including persistent challenges for current multimodal models, and, importantly, a lack of sensitivity to semantic distinctions in natural language descriptions. This analysis represents an initial step toward systematically studying semantic sensitivity in multimodal music models and highlights the value of MusicSem as a benchmark for this purpose. To support reproducibility, the MusicSem website and codebase provide complete instructions for reproducing all experiments reported in this paper.

\paragraph{\textbf{Organization of this Paper.}}
This paper is organized as follows. Section~\ref{sec:relatedwork} focuses on related work by presenting existing language--audio datasets for multimodal music representation learning. Section~\ref{sec:musicalsemantics} formalizes the notion of musical semantics, introduces our taxonomy, and analyzes the limited semantic awareness of models trained on existing datasets. Section~\ref{sec:musicsem} presents the MusicSem dataset, detailing the construction pipeline, key statistics and characteristics, and ethical considerations. Section~\ref{sec:evaluation} presents and discusses our experimental evaluation of multimodal models for retrieval and generation using MusicSem. Finally, Section~\ref{sec:conclusion} concludes the paper and outlines directions for future work.
\section{Related Work}
\label{sec:relatedwork}

In this section, we review existing language--audio datasets for multimodal music representation learning. These datasets differ substantially in the source of their textual annotations and the types of information they capture. We focus on those most relevant to our work and refer readers to the survey by \citet{christodoulou_multimodal_2024} for a comprehensive overview.

\subsection{Human-Annotated Language--Audio Datasets}

\begin{table}[t]
\caption{Statistics of existing language--audio music datasets that are most comparable to our setting. \textit{L--A Pairs} denotes the number of language--audio pairs, while \textit{Annotation} indicates the origin of the textual annotations describing the music.}
\label{tab:sota_dataset_stats}
\resizebox{\linewidth}{!}{
\begin{tabular}{l  cccc}
\toprule
\textbf{Dataset Name} & \textbf{Year} &\textbf{L-A Pairs} & \textbf{Annotation} & \textbf{Base Dataset} \\ 
\midrule 
MusicNet \citep{thickstun_invariances_2018} & 2018& 330 & Human& - \\
Song Describer \citep{manco_describer_2023}  & 2023 & 1,106 & Human & -\\ 
YouTube8M-MusicTextClips \citep{mckee_language_2023} &2023 &  4,169 & Human & - \\
MusicCaps \citep{agostinelli_musiclm_2023} & 2023 & 5,521 &Human & - \\
\midrule
MuLaMCap \citep{huang_noise2music_2023} & 2023 & 6,800,000& LLM & AudioSet\\ 
LP-MusicCaps \citep{doh_lpmusiccaps_2023} & 2023 & 2,000,000 & LLM & MusicCaps, Magnatagtune, \& Million Song Dataset \\ 
Text2Music~\citep{schneider_mousai_2024} &2024 &  50,000 & LLM & Spotify \\ 
FUTGA \citep{wu_futga_2024} & 2024 & 51, 800 &LLM& MusicCaps \& Song Describer\\ 
MusicBench \citep{melechovsky_mustango_2024} & 2024&53,168& LLM & MusicCaps \\ 
JamendoMaxCaps \citep{roy_jamendomaxcaps_2025} & 2025 & 200,000 & LLM & Jamendo \\ 
\bottomrule
\end{tabular}}
\end{table}

Several language--audio datasets have been introduced in recent years. However, as summarized in Table~\ref{tab:sota_dataset_stats}, only a limited number of them pair music audio with human-written textual annotations. Among them, \text{MusicCaps} is one of the most commonly adopted benchmarks for music--language alignment~\citep{agostinelli_musiclm_2023}. It contains 5,521 audio clips annotated by professional musicians, with captions that emphasize descriptive musical attributes such as instrumentation, genre, tempo, and stylistic characteristics. Similarly, \text{YouTube8M-MusicTextClips}~\citep{mckee_language_2023} provides 4,169 language--audio pairs, where captions are written by hired annotators following predefined guidelines.

Crowd-sourced annotation has also been explored. \text{Song Describer}~\citep{manco_describer_2023} extends a subset of the \text{Jamendo} dataset~\citep{bogdanov_mtg_2019} with 1,106 crowd-written descriptions. In a complementary but distinct line of work, \text{MusicNet} provides human-annotated, note-level labels verified by trained musicians and aligned with audio recordings for 330 freely licensed classical music recordings~\citep{thicstun_musicnet_2017}. While MusicNet offers rich symbolic and structural annotations, it does not include actual natural language descriptions of music, and is therefore primarily used for symbolic music understanding rather than multimodal language--audio representation learning.

Overall, these datasets provide high-quality textual annotations, but their natural language descriptions are typically constrained by annotation instructions or musical expertise, resulting in a focus on acoustic, technical, or stylistic aspects of music. Moreover, their scale is limited to at most a few thousand language--audio pairs.

\subsection{LLM--Annotated Language--Audio Datasets}

To scale language--audio supervision, recent work has increasingly relied on large language models (LLMs) to generate or augment textual annotations~\citep{naveed2025comprehensive}. The \text{MuLaMCap} dataset~\citep{huang_noise2music_2023} combines \text{MusicCaps} with LLM-generated descriptions of 150K popular songs, producing a dataset of 6.8M language--audio pairs. \text{LP-MusicCaps}~\citep{doh_lpmusiccaps_2023} merges multiple music datasets, including \text{MusicCaps}, \text{MagnaTagATune}~\citep{law_evaluation_2009}, and the \text{Million Song Dataset}~\citep{mcfee_million_2012}, and uses LLMs to generate 2.2M sentence-like captions paired with 500K audio samples.

Other works focus on augmenting existing captions with automatically extracted musical structure. \text{MusicBench}~\citep{melechovsky_mustango_2024} enriches \text{MusicCaps} with downbeat, chord, key, and tempo information derived from signal processing algorithms. The dataset used to train \text{FUTGA}~\citep{wu_futga_2024} follows a similar strategy, prompting an LLM to augment captions from \text{MusicCaps} and \text{Song Describer} with structural descriptors. \text{JamendoMaxCaps}~\citep{roy_jamendomaxcaps_2025} generates captions for 200K \text{Jamendo} tracks using a music captioning model, while \text{Text2Music}~\citep{schneider_mousai_2024} scrapes playlist metadata from Spotify and reformulates them into sentence-level descriptions using an LLM.

Although these datasets substantially increase scale, they rely heavily on synthetic text generation. They often lack explicit discussion of hallucination mitigation, annotation reliability, or how well the generated text reflects real-world listening experiences, despite evidence that musical meaning and perception are highly subjective and context-dependent \citep{sloboda2001psychological,levy_learning_2008,choi_prediction_2020,epure2020modeling}.

\subsection{Other Related Datasets}

For completeness, we note the existence of additional datasets that consist primarily of processed versions of the ones discussed above. In the context of generative retrieval and musical question answering, \text{MusicQA}~\citep{liu_music_2024} reformulates captions from \text{MusicCaps} and \text{MagnaTagATune} into approximately 4,500 question--answer pairs using an LLM. Similarly, \text{LLaRK}~\citep{gardner_llark_2024} constructs over 1.2M language--audio pairs by aggregating multiple music datasets, including \text{MusicCaps}, \text{YouTube8M}, \text{MusicNet}, \text{Jamendo}, and \text{MagnaTagATune}, as well as \text{FMA}~\citep{fma}, which associates audio files with metadata and free-form textual content such as artist biographies.

Other related datasets are also worth mentioning, although they fall partially outside the scope of this work. For instance, \text{MMAU}~\citep{sakshi_mmau_2024} curates 10K general audio and music question--answer pairs to evaluate diverse music understanding and reasoning capabilities. Several datasets also draw from Reddit, as we do for \text{MusicSem}, but with different objectives. \text{Tip-of-the-Tongue}~\citep{bhargav_tot_2023} collects posts from \texttt{r/TipOfMyTongue} to study search scenarios in which users have previously experienced an item (e.g., a song or a movie) but cannot recall a reliable identifier. \citet{vaselovsky_imagine_2021} collect over 536K song--artist pairs from Reddit to analyze music sharing behavior across communities. Beyond Reddit, the \text{Million Tweet Dataset}~\citep{hauger_million} includes over one million music-related tweets to study popularity trends and cultural dynamics.

Nonetheless, while these datasets highlight the value of large-scale textual data for studying music-related behavior, they are not designed as paired language--audio resources for training and evaluating multimodal music representation learning models.

\section{Capturing Musical Semantics in Multimodal Representation Learning}
\label{sec:musicalsemantics}

We now introduce the key concept of musical semantics in more detail. We first present our taxonomy and discuss the importance of musical semantics in Section~\ref{sec:taxonomy}. We then examine the relative lack of semantic awareness in many current multimodal models in Section~\ref{sec:insensitivity}.

\subsection{A Taxonomy of Musical Semantics}
\label{sec:taxonomy}

Throughout this paper, we use the term \textit{musical semantics} to refer to the aspects of meaning conveyed in textual descriptions of music. These aspects go beyond acoustic properties of music and contextualize the overall listening experience. They capture how listeners interpret, experience, and situate music in relation to emotions, activities, social contexts, and prior knowledge \citep{levy_learning_2008,nam_deep_2019,choi_prediction_2020}.

\begin{table}[t]
 \centering
\caption{Examples of caption elements illustrating the five different categories in our proposed taxonomy of musical semantics.}
 \label{tab:categories}
\resizebox{\linewidth}{!}{
    \begin{tabular}{c|l}
        \toprule
        \textbf{Category} & \textbf{Example of textual description of music} \\
        \midrule
        Descriptive & \textit{``I like the high pass filter on the vocals in the chorus, really makes harmonies pop.''} \\ 
        Contextual & \textit{``Sabrina Carpenter's Espresso is just a mix of old Ariana Grande and 2018 Dua Lipa.''} \\ 
        Situational & \textit{``I listened to this song on the way to quitting my exhausting corporate job.''} \\ 
        Atmospheric & \textit{``This song makes me feel like a manic pixie dream girl in a bougie coffeeshop.''} \\ 
        Metadata  & \textit{``This deluxe edition of this song was released in 2013 and it has three bonus hiphop tracks.''} \\ 
        \bottomrule
    \end{tabular}
    }
\end{table}

We propose to organize these elements into five categories, which together form our \textit{taxonomy of musical semantics}. Our objective is to provide a structured view of the different types of meaning listeners commonly express when describing music. We distinguish the following five categories, with illustrative examples for each presented in Table~\ref{tab:categories}:
\begin{enumerate}
    \item \textbf{Descriptive} semantics, which characterize intrinsic musical attributes such as instrumentation, genre, tempo, or style;
    \item \textbf{Contextual} semantics, which relate a song to other music through similarity, comparison, or co-listening patterns;
    \item \textbf{Situational} semantics, which describe activities, settings, or environments in which a song is typically listened to;
    \item \textbf{Atmospheric} semantics, which express emotions, moods, or other affective and expressive qualities evoked by the music;
    \item \textbf{Metadata-based} semantics, which provide technical, historical, or background information about a song or its artist.
\end{enumerate} For more details on how these categories were selected, please see Appendix~\ref{ap:sem_cat}.

Modeling these semantic distinctions is critical for multimodal music representation learning. To illustrate this point, we consider again the example introduced in the introduction with the following two natural language descriptions: \textit{``This song is a ballad. It contains guitar, male vocals, and a piano. It sounds like something I would listen to at church.''} versus \textit{``This song is a ballad. It contains guitar, male vocals, and a piano. It sounds like something I would listen to during a psychedelic experience.''} Both descriptions specify identical descriptive musical attributes. However, the \textit{situational} context would alter a listener's expectations regarding the audio that should be generated by a text-to-music model, as well as the audio that should be retrieved by a multimodal retrieval system given these prompts.
This example highlights how shifts in musical semantics, even when subtle, can lead to qualitatively different interpretations, underscoring the importance of explicitly modeling semantic variations in language--audio representations.

\subsection{Limited Sensitivity to Semantic Context in Current Multimodal Models}

To motivate the introduction of MusicSem in the next section of this paper, we conduct a preliminary exploratory study examining the sensitivity of several widely used multimodal music representation learning models to variations in musical semantics.

\paragraph{\textbf{Setting.}}
We design the following experiment. Given a language--audio pair $(t_i, a_i)$ from a dataset, we construct a counterfactual annotation $\tilde{t}^x_i$ by modifying the original text according to a specific semantic category $x$, for example replacing \textit{``while at church''} with \textit{``during a psychedelic experience.''} We sample 50 language--audio pairs from the \text{MusicCaps} dataset \citep{agostinelli_musiclm_2023} and have trained musicians manually construct counterfactual annotations for each semantic category present in the captions. The complete set of counterfactual examples derived from MusicCaps is publicly released at: \url{https://github.com/Rsalganik1123/MusicSem/blob/main/data/counterfactual_examples/all_counterfac.csv}.

To assess the sensitivity of text-to-music generative models to semantic shifts, we define the following metric:
\begin{equation}
G^x = \frac{1}{n} \sum_{i = 1}^n \left[1 - \text{cosine}(f_i, \tilde{f}^x_i)\right],
\label{eq:g}
\end{equation}
where $n$ denotes the number of language--audio pairs, and $f_i = \mathcal{M}(t_i)$ and $\tilde{f}^x_i = \mathcal{M}(\tilde{t}^x_i)$ correspond to the model outputs of the text-to-music generative model $\mathcal{M}$ given the original and counterfactual text inputs, respectively.

To evaluate the sensitivity of text-to-music retrieval models, we define:
\begin{equation}
R^x@k = \frac{1}{n} \sum_{i=1}^n \left[1 - \frac{|A_i \cap \tilde{A}^x_i|}{|A_i|} \right],
\end{equation}
where $A_i = \mathcal{M}(t_i)$ and $\tilde{A}^x_i = \mathcal{M}(\tilde{t}^x_i)$ denote the sets of top-$k$ audio candidates retrieved by the model given the original and counterfactual textual inputs, respectively.

Table~\ref{tab:sensitivity_t2m} reports the sensitivity to semantic shifts of several popular text-to-music generative and retrieval models from recent literature. All evaluated models were trained on existing language--audio datasets reviewed in Section~\ref{sec:relatedwork} and, in some cases, on additional proprietary datasets that are not publicly available. We refer readers to the corresponding references and our Appendix~\ref{app:experimentalsettings} for implementation and training details of each model.

\paragraph{\textbf{Results.}}
\label{sec:insensitivity}
The results in Table~\ref{tab:sensitivity_t2m} indicate that these models exhibit substantially greater sensitivity to changes in descriptive musical attributes than to shifts in atmospheric, situational, contextual, or metadata-related semantics. This pattern highlights a relative lack of semantic awareness in current textual conditioning mechanisms and suggests that these models struggle to capture the expectations implied by user intent beyond surface-level musical descriptors. Nonetheless, as argued in Section~\ref{sec:taxonomy}, modeling such semantic variations remains crucial for faithful and user-aligned multimodal music generation and retrieval. These findings motivate the need for a semantics-aware dataset that better reflects the diversity of musical semantics expressed in natural language, in order to support the training and evaluation of multimodal models.

\begin{table}[t]
\centering
    \caption{Semantic sensitivity analysis of generative (top) and retrieval (bottom) models. Best performance is in bold. Superscripts $^d$, $^a$, $^s$, $^m$, and $^c$ denote descriptive, atmospheric, situational, metadata, and contextual, respectively.}
        \label{tab:sensitivity_t2m}
    \resizebox{0.8\linewidth}{!}{
    \begin{tabular}{l|ccccc}
    \toprule 
         \textbf{Text-to-Audio Generative Model} & $G^d$&$G^a$&$G^s$  &$G^m$ & $G^c$ \\
          \midrule
          AudioLDM2~\citep{liu_audioldm2_2024} & {0.68} &0.37 &  0.35 & 0.40 & 0.34 \\ 
          MusicLM~\citep{agostinelli_musiclm_2023} & 0.50 & 0.36 & {0.42} & 0.39 & 0.35\\  
          Mustango~\citep{melechovsky_mustango_2024} & 0.62 & 0.27 & 0.25 & 0.26 & 0.32  \\ 
          MusicGen~\citep{copet_simple_2023} & 0.57 & {0.47} & 0.39 & {0.47} & {0.52} \\ 
          Stable Audio~\citep{evans_fast_2024} & \textbf{0.72} & \textbf{0.67} & \textbf{0.68} & \textbf{0.70} & \textbf{0.74} \\ 
          % Mureka\footnotemark & - & - & - & - & - \\ 
         \midrule
         \midrule       
         \textbf{Text-to-Audio Retrieval Model} ($k =10) $ & $R^d$ & $R^a$ & $R^s$ & $R^m$ & $R^c$ \\
          \midrule
          LARP~\citep{salganik_larp_2024} & \textbf{0.98} & 0.17 & 0.06 & 0.0 & \textbf{0.56}\\ 
          CLAP~\citep{wu_largescale_2023} & {0.95} & {0.52} & {0.35} & {0.42} & 0.52 \\ 
           ImageBind~\citep{girdhar_imagebine_2023} & 0.84 & 0.39 & {0.35} & 0.38 & 0.41 \\
            CLaMP3~\citep{wu_clamp3_2025} & 0.92 & \textbf{0.58} & \textbf{0.49} & \textbf{0.62} & {0.55} \\ 
         \midrule
    \end{tabular}}
\end{table}

\section{MusicSem: A Language--Audio Dataset of Natural Music Descriptions on Reddit Capturing Musical Semantics}
\label{sec:musicsem}

To address this lack of semantic sensitivity, we introduce \textit{MusicSem}, a novel dataset of language--audio music description pairs extracted from organic musical discourse on the social media platform Reddit\footnote{\url{https://www.reddit.com/}}. MusicSem is designed to capture richer and more nuanced musical semantics, supporting the study, training, and evaluation of multimodal music representation learning models. Its construction requires multiple processing stages, including identifying, extracting, structuring, and validating semantic content from online discourse, combining LLM-assisted extraction and summarization with human verification.

We detail the complete dataset construction process in Section~\ref{sec:datasetconstruction}. Section~\ref{sec:datasetanalysis} provides a descriptive analysis of the final MusicSem dataset, while Section~\ref{sec:datasetdiscussion} examines key characteristics and implications related to music genre and cultural representativeness, as well as the presence of subjective content. Finally, Section~\ref{sec:ethics} discusses the key safeguards adopted to ensure ethical and responsible data processing and release.

\subsection{Dataset Construction}
\label{sec:datasetconstruction}

\paragraph{Reddit Thread Selection.}

Reddit is a large-scale online discussion platform where users engage in topic-centered communities, known as \textit{subreddits}, to share content and exchange opinions. Reddit discussions are organized into threads, each initiated by a post and followed by a sequence of user comments, often forming detailed and context-rich conversations. They provide a natural source of user-generated musical discourse, reflecting how listeners describe, interpret, and contextualize music  \citep{medvedev2017anatomy,reddit2}.

To construct MusicSem, we extract music-related textual descriptions from five popular English-language subreddits that feature sustained user discussions and cover a broad range of musical styles and listening practices:
\begin{enumerate}
    \item \texttt{r/electronicmusic}, which focuses on electronic music genres, with discussions ranging from production and stylistic characterization to listening contexts;
    \item \texttt{r/popheads}, a community centered on pop music, where users frequently discuss new releases, artists, and their personal listening experiences;
    \item \texttt{r/progrockmusic}, which emphasizes progressive rock and related genres, often featuring detailed discussions of musical structure, composition, and artist influence;
    \item \texttt{r/musicsuggestions}, a recommendation-oriented subreddit in which users describe musical preferences, moods, or situations to solicit tailored listening suggestions;
    \item \texttt{r/LetsTalkMusic}, a general music discussion forum that encourages reflective conversations about music, spanning various genres, eras, and personal interpretations.
\end{enumerate}

\begin{figure}[ht]
    \centering
    \includegraphics[width=\linewidth]{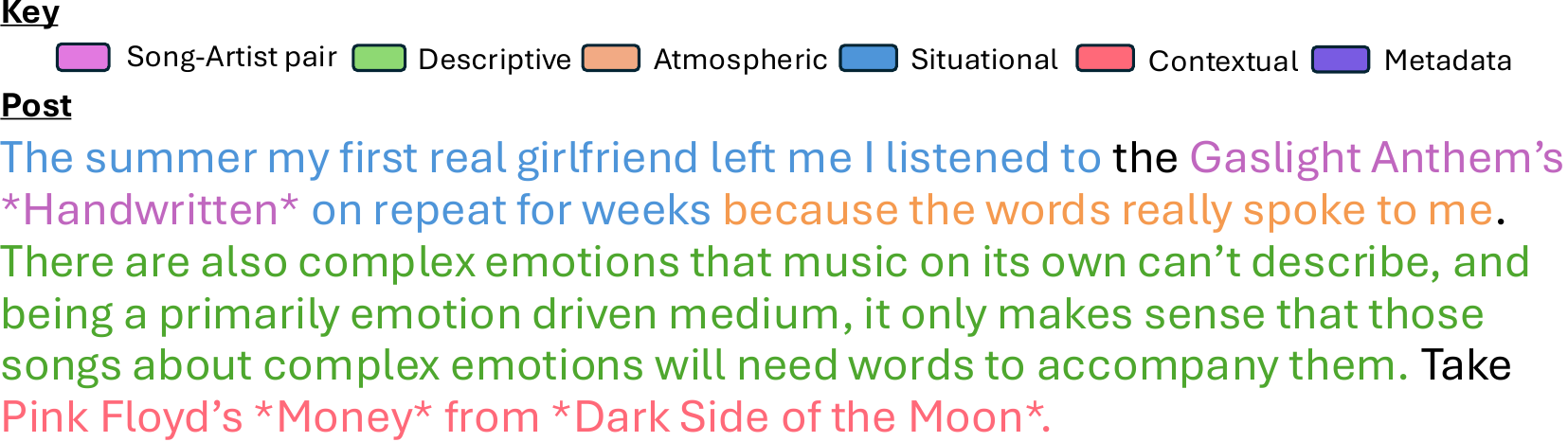}
\caption{Example of semantic content extracted from a Reddit post in MusicSem. The figure highlights how a single description can express a variety of different musical semantics, corresponding to the five categories defined in our taxonomy.}
    \label{fig:motivation}
\end{figure}

\begin{figure}[ht]
    \centering
    \includegraphics[width=\linewidth]{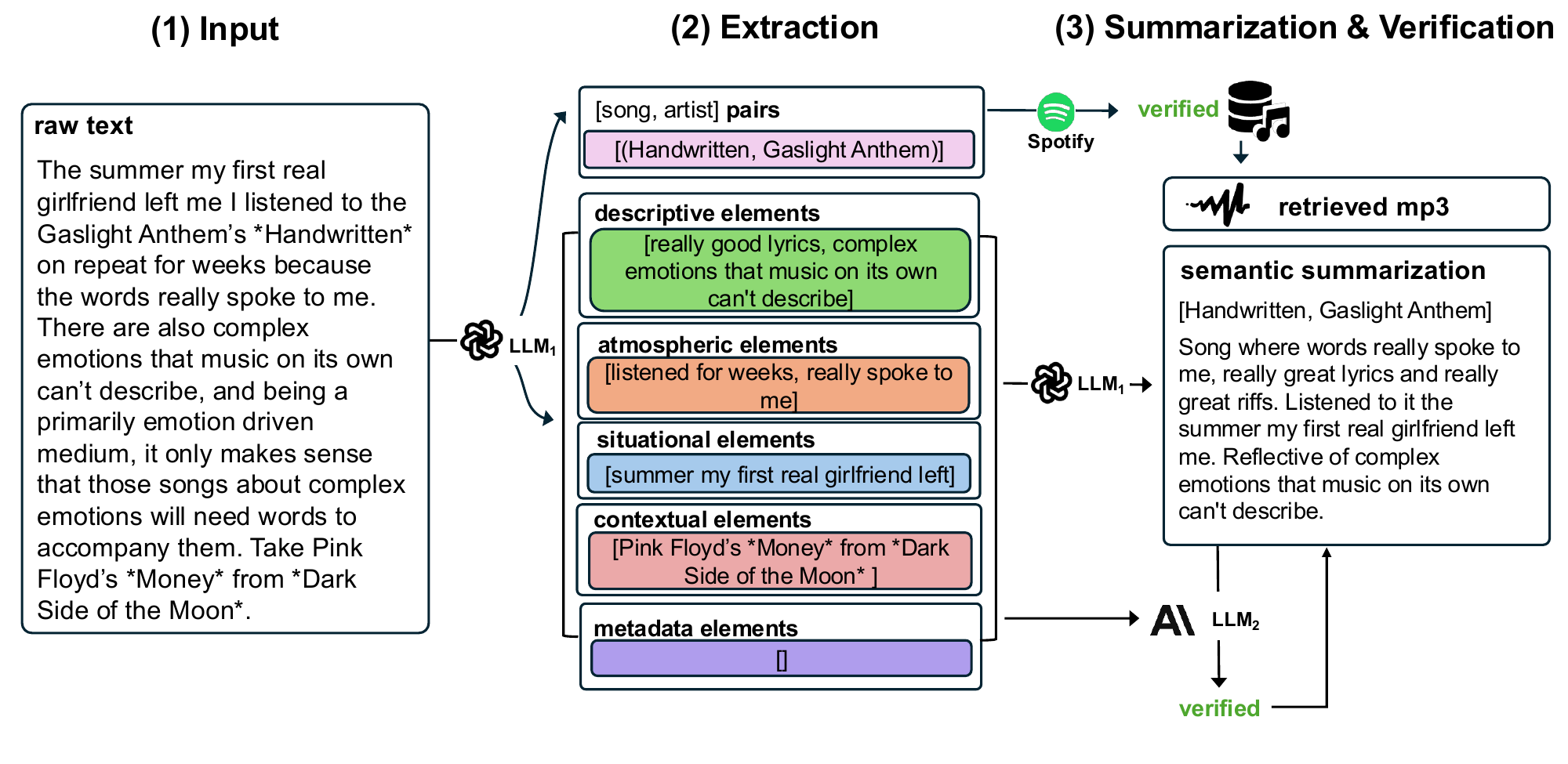}
\caption{Overview of the extraction and verification pipeline used to construct MusicSem. After selecting the source Reddit threads, the dataset construction proceeds in two main stages: an extraction step that identifies candidate semantic content from the textual elements of each thread, and a summarization and verification step that reformulates the extracted content into sentence-like semantic annotations, verifies song--artist associations, and checks the plausibility of the extracted semantic information.}
    \label{fig:extract}
\end{figure}

We collect data using the Pushshift API\footnote{\url{https://github.com/pushshift/api}}, covering discussions from 2008 to 2022.
Then, the dataset construction pipeline includes various extraction, summarization, and verification steps, as described below and illustrated in Figures~\ref{fig:motivation} and~\ref{fig:extract}.

\paragraph{Information Extraction.}
This phase aims to transform raw Reddit posts into structured representations that associate mentioned songs with their corresponding artists and semantic attributes. We concatenate the title and body of each post into a single prompt and submit it to a large language model for semantic extraction. We use GPT-4o (2024-08-06)~\citep{gpt4o} as the extraction model. To extract musical semantics, we design a prompt that instructs the model to identify \textit{(song, artist)} pairs and assign them semantic content across the categories defined in Section~\ref{sec:taxonomy}. Inspired by prior work on knowledge extraction with language models~\citep{jiang-etal-2020-know}, the prompt is manually crafted using illustrative examples and iteratively refined through interaction with the model. The full set of prompts used at this stage is provided in Appendix~\ref{app:construction}.

\paragraph{Information Summarization and Song--Artist Verification.}

First, we format the extracted semantic content into sentence-level annotations and verify the correctness of \textit{(song, artist)} associations. Following practices adopted in existing language--audio datasets \citep{agostinelli_musiclm_2023,manco_describer_2023,mckee_language_2023}, we use LLMs to rephrase extracted semantic tags into natural language sentences. To mitigate incorrect song--artist associations, we verify each extracted \textit{(song, artist)} pair by measuring the character-level overlap between the extracted names and the original post text, after lowercasing all strings. Pairs with less than 75\% overlap are discarded. We then query the Spotify API\footnote{\url{https://developer.spotify.com/documentation/web-api}} to retrieve a unique identifier for each remaining pair. In cases where multiple candidate entries are returned, we apply the same overlap-based filtering strategy to resolve ambiguities.

\paragraph{Audio Retrieval.}
Once a unique identifier is obtained, we retrieve the corresponding audio using \textit{spotdl}\footnote{\url{https://github.com/spotDL/spotify-downloader}}. If no audio file can be retrieved, the associated \textit{(song, artist)} pair is removed from the dataset. After collecting the validated \textit{(song, artist)} pairs, their associated audio files, and the extracted semantic tags, we use GPT-4o to rephrase the semantic content into sentence-like annotations. 
To respect copyright constraints, we release the unique identifiers associated with each song and our complete data construction pipeline, rather than the audio files themselves.

\paragraph{Verification for Accuracy, Faithfulness, and Hallucination Mitigation.}

Finally, to verify the faithfulness of the generated annotations and associations, we employ an independent verification model, Claude Sonnet 3.7~\citep{claude3.7}, which compares the original extracted semantic tags with the rephrased annotations. The verification model is prompted to produce a binary decision indicating whether the annotation is consistent or hallucinated, and entries flagged as hallucinated are removed. The complete corresponding prompt for these verification steps is provided in Appendix~\ref{app:construction}. 

We further complement these LLM-assisted verification steps with a detailed series of manual checks conducted by our team, including two professionally trained musicians, to ensure accuracy, faithfulness, and effective hallucination mitigation in the final dataset.

\paragraph{Final Dataset.} 
The final MusicSem dataset consists of 32,493 language–audio pairs. Each entry contains the following fields:
\begin{itemize}
\item \texttt{unique\_id}: unique identifier used by Spotify to identify the track;
\item \texttt{thread}: name of the subreddit from which the Reddit raw post is extracted;
\item \texttt{spotify\_link}: URL to the Spotify web player for the track;
\item \texttt{song}: song title;
\item \texttt{artist}: artist name;
\item \texttt{raw\_text}: concatenation of the post title and body;
\item \texttt{prompt}: LLM-generated summary of semantic extractions from the raw post;
\item \texttt{descriptive}: list of strings capturing descriptive elements;
\item \texttt{contextual}: list of strings capturing contextual elements;
\item \texttt{situational}: list of strings capturing situational elements;
\item \texttt{atmospheric}: list of strings capturing atmospheric elements;
\item \texttt{metadata}: list of strings capturing metadata elements;
\item \texttt{pairs}: list of tuples containing all song–artist pairs mentioned in the post.
\end{itemize}

To facilitate meaningful evaluation, we curate a human-validated test set of 480 entries. This test set is made available to reviewers at \url{https://tinyurl.com/3n8je74z}
 and will be removed upon publication. It will remain unpublished thereafter to support the development of a future leaderboard. All remaining entries constitute the public training set.

MusicSem is publicly released under the MIT License and is available on Hugging Face:
\url{https://huggingface.co/datasets/AMSRNA/MusicSem}. The complete source code used to construct MusicSem is available on GitHub:
\url{https://github.com/Rsalganik1123/MusicSem}. This repository also provides instructions for collecting data from additional music-related subreddits using the Pushshift API.  All resources are also directly accessible via the accompanying MusicSem website (see Figure~\ref{fig:website} in the introduction), which includes documentation and visualizations.

\subsection{Descriptive Analysis of the Dataset}
\label{sec:datasetanalysis}

We now present an exploratory descriptive analysis of the MusicSem dataset.

\paragraph{Musical Semantics Diversity.}
First, we analyze the diversity of musical semantics captured by MusicSem in comparison with existing language–audio datasets. Table~\ref{tab:semantic_analysis} reports the proportion of entries containing each of the five semantic categories for MusicSem and the two most widely used human-annotated language–audio datasets discussed in Section~\ref{sec:relatedwork}: MusicCaps \citep{agostinelli_musiclm_2023} and Song Describer \citep{manco_describer_2023}.

As shown in Table~\ref{tab:semantic_analysis}, MusicSem consistently exhibits broader coverage across all semantic categories. For example, 77\% of entries include contextual tags, compared to 6\% for MusicCaps and 8\% for Song Describer. These results indicate substantially richer semantic content.
In addition, MusicSem constitutes a significantly larger dataset than MusicCaps and Song Describer, which contain only 5,521 and 1,100 entries, respectively. It also features a richer vocabulary, with a higher number of unique tokens and music genre references.

\begin{table}[t]
    \centering
\caption{General statistics (top) and coverage by music semantic category (bottom) for MusicSem and two other canonical language–audio music datasets.}
    \label{tab:semantic_analysis}
    \resizebox{0.85\textwidth}{!}{\begin{tabular}{c|ccc}
        \toprule
                    \multirow{2}{*}{\textbf{Statistics}} & \textbf{MusicCaps}  & \textbf{Song Describer} & \textbf{MusicSem} \\
                     & \citep{agostinelli_musiclm_2023} &  \citep{manco_describer_2023} & (ours) \\
    \midrule            Number of entries & 5,521 & 1,100 & \textbf{32,493} \\
    Number of distinct words & 6,245 & 2,824 & \textbf{22,738} \\
    Number of distinct music genres & 267 & 152 & \textbf{493} \\
    \midrule
    \midrule
                    \multirow{2}{*}{\textbf{Category of Musical Semantics}} & \textbf{MusicCaps}  & \textbf{Song Describer} & \textbf{MusicSem} \\
                     & \citep{agostinelli_musiclm_2023} &  \citep{manco_describer_2023} & (ours) \\
        \midrule
        Descriptive & \textbf{100\%} & 94\% & \textbf{100\%} \\
        Contextual & 6\% & 8\% & \textbf{77\%} \\
        Situational & 41\% & 16\% & \textbf{48\%} \\
        Atmospheric & 57\% & 33\% & \textbf{64\%} \\
        Metadata & 28\% & 6\% & \textbf{64\%} \\
    \bottomrule
    \end{tabular}}
\end{table}

\paragraph{Personalization and Contextualization.}
While MusicSem contains 32,493 entries, these correspond to 11,842 unique songs and 4,430 unique artists, reflecting the fact that many songs are discussed across multiple posts. We view this characteristic as an advantage of the dataset. Beyond semantic diversity, MusicSem captures two distinctive properties that are largely absent from existing  datasets: \textit{personalization} and \textit{contextualization}.

\begin{figure}[t]
    \centering
    \includegraphics[width=\textwidth]{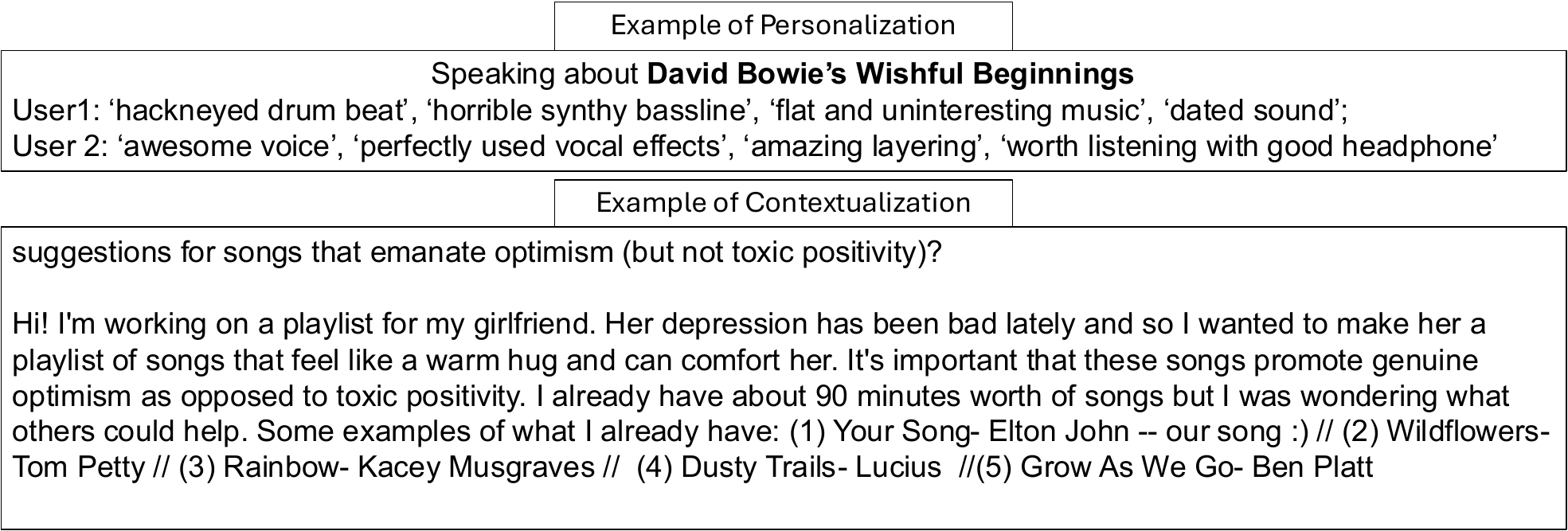}
    \caption{An example of personalization and contextualization on Reddit. }
    \label{fig:dataset_case_study}
\end{figure}

Regarding personalization, each song is discussed in an average of 2.98 distinct posts, resulting in multiple, potentially divergent semantic descriptions of the same musical piece. This multiplicity reflects individual listener perspectives and use cases, and enables the study of personalized or user-dependent interpretations of music. An example is shown in Figure~\ref{fig:dataset_case_study}, where different users associate distinct semantic attributes with the same song.

In terms of contextualization, posts in MusicSem mention an average of 10.51 songs, which are often grouped under shared themes, moods, or situational contexts (such as positivity or relaxation). Such co-occurrences provide explicit signals of semantic relatedness between songs, grounded in user-defined contexts rather than editorial taxonomies. This structure highlights the importance of modeling inter-song associations and latent user intent in music understanding and retrieval tasks.

\begin{figure}[t]
    \centering
    \begin{subfigure}{0.6\textwidth}
        \centering
        \includegraphics[width=\linewidth]{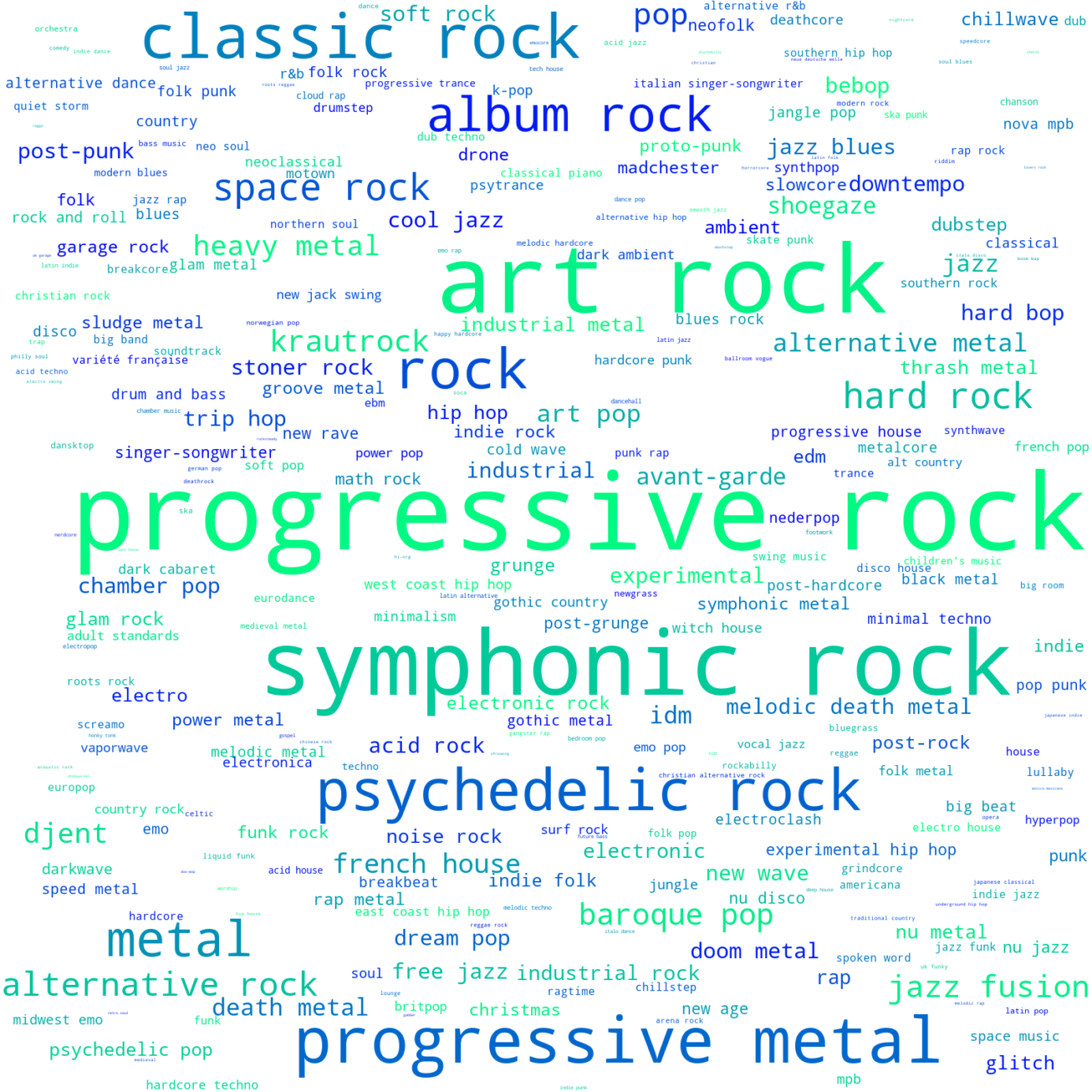}
        \caption{}   
    \end{subfigure}%
    \hfill%
    \begin{subfigure}{0.33\textwidth}
        \centering
        \includegraphics[width=\linewidth]{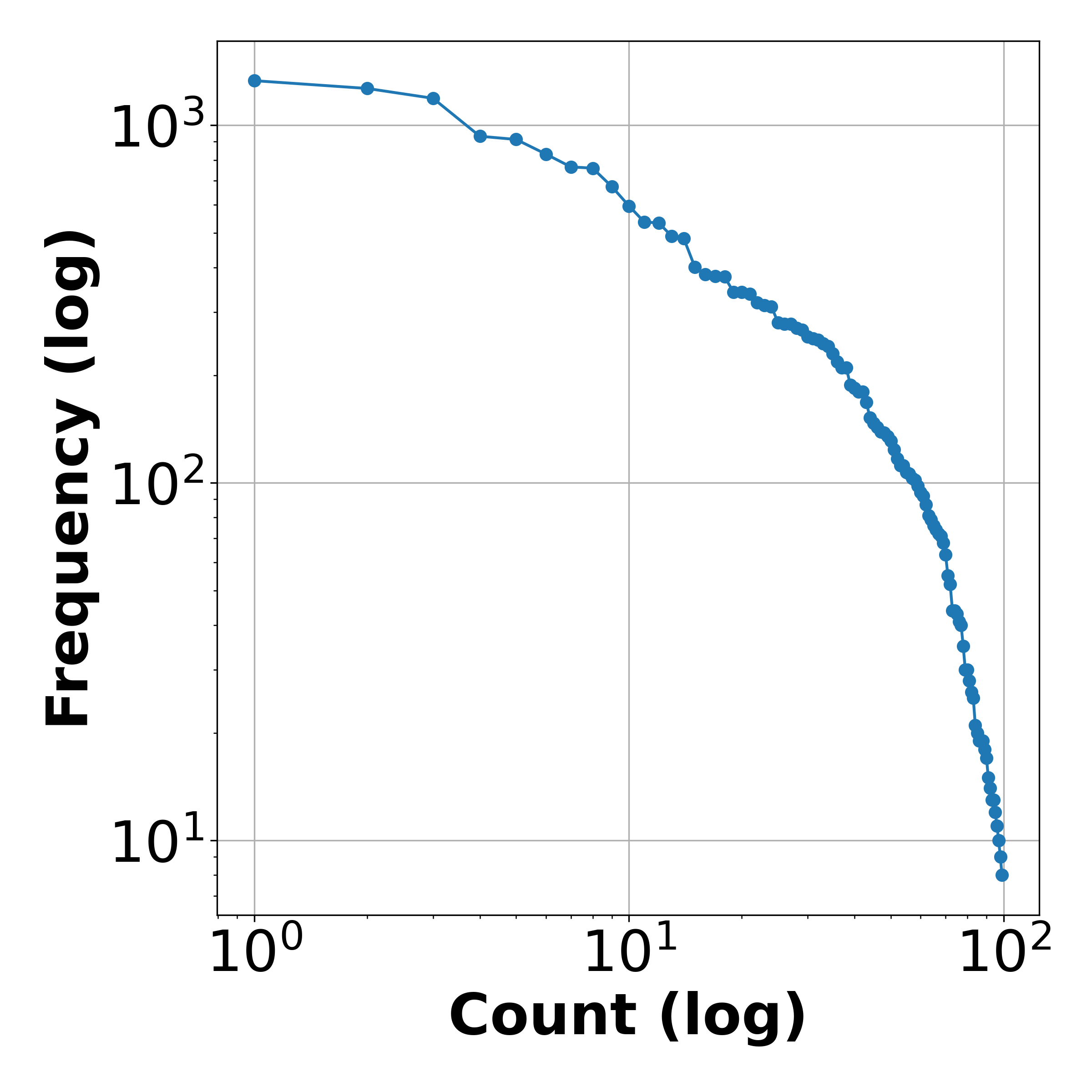}
        \caption{}
        \vspace{0.8em}
        \includegraphics[width=\linewidth]{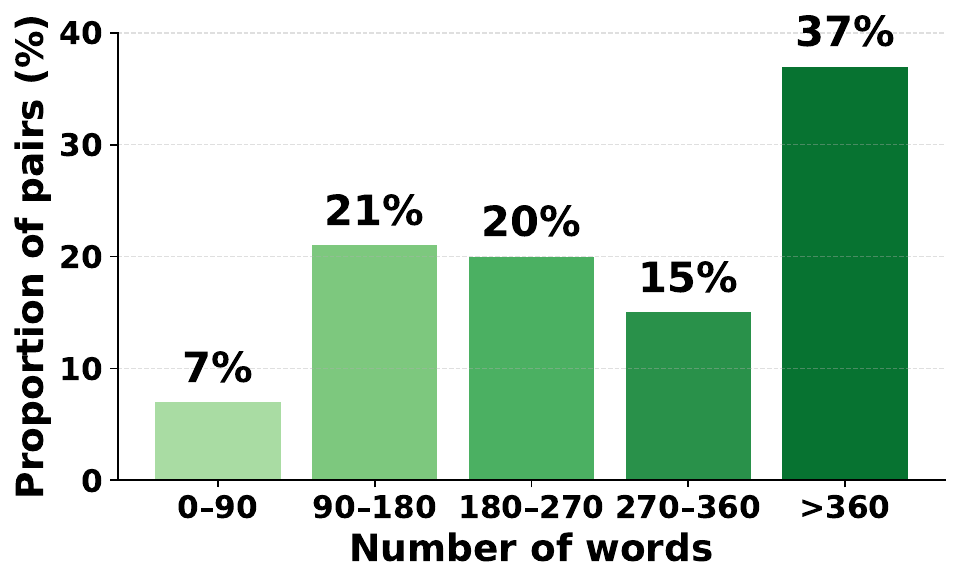}
        \caption{}
    \end{subfigure}

    \caption{Visualizations for MusicSem. (a) Music genre distribution visualized as a word cloud, where larger font size indicates higher frequency. (b) Popularity distribution of songs. (c) Distribution of the number of words per language--audio pair.}
    \label{fig:dataset-visu}
\end{figure}

\paragraph{Visualizations.}
Figure~\ref{fig:dataset-visu} presents additional visualizations illustrating several aspects of MusicSem. Figure~\ref{fig:dataset-visu}(a) shows the distribution of music genres in MusicSem, with strong representation of rock, metal, electronic, and pop music. This distribution is expected given the five subreddits selected for dataset construction. We further discuss the resulting music genre bias and its practical implications in Section~\ref{sec:datasetdiscussion}.

Figure~\ref{fig:dataset-visu}(b) depicts the distribution of song occurrences across posts. The dataset follows a power-law distribution, in which a small number of songs are mentioned frequently while many others appear only rarely. This pattern aligns with well-documented popularity biases in music datasets, which often lead to substantial disparities between mainstream and niche music representation~\citep{salganik_mitigating_2024}.

Finally, Figure~\ref{fig:dataset-visu}(c) reports the distribution of raw post lengths. The dataset is skewed toward longer discussions, with most posts exceeding 360 characters. This contributes to the rich vocabulary and high density of semantic content observed in MusicSem.

%This selection bias is broadly addressed in the limitations portion of our dataset. In this version of our dataset we focus on finding semantically rich musical discourse on Reddit without specifically considering genre coverage. In the future we hope to continue expanding the scope of the dataset to include a broader variety of genres. 

\subsection{Discussion of Dataset Characteristics and Implications}
\label{sec:datasetdiscussion}

MusicSem is designed to capture a broad spectrum of musical semantics as they naturally emerge from listener discourse, reflecting nuanced, human-centered descriptions of music. While this focus enables rich semantic analysis, several key characteristics of the dataset should be carefully considered when using it. We discuss these considerations below.

\paragraph{Music Genre Representativeness.}
As shown in Figure~\ref{fig:dataset-visu}(a), certain music genres are overrepresented in MusicSem. This imbalance is a direct consequence of the subreddit selection process. For example, sourcing data from \texttt{r/progrockmusic} naturally results in a higher prevalence of progressive rock discussions. In its current version, MusicSem prioritizes semantically rich musical discourse over balanced genre coverage. Consequently, it should not be interpreted as genre-representative of music listening behavior at large.
Future users should be cautious when conducting genre-specific analyses, as some genres are underrepresented. That said, as explained in Section~\ref{sec:datasetconstruction}, our public repository provides detailed instructions for collecting data from additional music-related subreddits, enabling users to augment the dataset with annotations from other communities and genres as needed.

\paragraph{Cultural Representativeness.}
Prior work has demonstrated that cultural background, language, socio-economic factors, and individual musicological experience strongly influence how music is perceived and described \citep{sordo2008quest,morrison2009cultural,lee2013kpop,epure2020modeling}. In this context, we acknowledge that the construction of MusicSem, like most datasets derived from online platforms, introduces inherent selection biases that affect its linguistic and semantic characteristics.
All textual data is collected from English-language subreddits, and Reddit’s user base is known to skew toward younger, digitally engaged, and predominantly Western populations \citep{barthel2016seven,epure2023human}. Moreover, users who actively participate in music-related discussions often hold strong opinions or are affiliated with niche communities \citep{medvedev2017anatomy,epure2023human}. These factors may influence both the content and framing of musical descriptions. We therefore encourage users to account for these properties when applying MusicSem to tasks involving user modeling or cross-cultural generalization.

\paragraph{Subjectivity and Content Noise.}
MusicSem relies on user-authored text, which is inherently subjective and may be informal or ambiguous. This variability constitutes a central strength of the dataset, capturing personal interpretations, emotions, and listening contexts. That said, it also introduces noise and inconsistencies in semantic descriptions. As such, MusicSem is not intended to provide authoritative annotations. Instead, it reflects the diversity of listener perspectives encountered in natural discourse.

In addition, the construction of MusicSem involves LLM-assisted extraction and processing. To mitigate potential errors, we incorporate rigorous faithfulness and hallucination in our pipeline, and our team manually reviewed the extracted content to verify (i) alignment between extracted songs and those explicitly mentioned in each thread, (ii) fidelity of musical attributes to the original user posts, and (iii) the absence of clear hallucinations. Despite these safeguards, some small degree of residual noise may remain, which should be taken into account when using the dataset.

\subsection{Ethical Considerations}
\label{sec:ethics}

MusicSem is constructed from publicly available online content and is intended for research use. We take ethical and legal considerations seriously and outline below the key measures adopted during dataset construction as ethical safeguards.

\paragraph{User Privacy and Anonymity.}
MusicSem is derived from Reddit threads that users voluntarily shared on a public platform designed around pseudonymous participation, where contributors are encouraged to conceal their real-world identities through self-selected identifiers. As a result, posts in MusicSem are authored under pseudonyms rather than identifiable personal names, substantially reducing the risk of linking content to specific individuals.

To further mitigate privacy risks, all user identifiers are removed prior to release. The dataset does not include usernames, user IDs, profile links, or other direct identifiers. In addition, two members of the author team manually reviewed all posts to remove any remaining explicit personal information (e.g., names, addresses, phone numbers, or email addresses). To the best of our knowledge, no such information remains in the dataset.

\paragraph{Release of Searchable Raw Reddit Posts.}
We acknowledge that some raw text from Reddit threads may be searchable through public search engines such as Google, as well as Reddit's own search functionality. For this reason, we carefully considered whether to exclude the raw post text from the release of MusicSem, and we remain open to considering it again should community standards or platform policies evolve in the future. 

However, at the time of release, we chose to preserve the original text, as all raw text that might be searchable was authored under pseudonymous identifiers. Retaining the raw text is important for maintaining the integrity of organic musical discourse and is essential for a range of research tasks, including semantic extraction, discourse analysis, and robustness evaluation of language models. We emphasize that the inclusion of raw text is intended solely to support research on music understanding and does not aim to increase the traceability of individual users beyond what is already possible on the original platform.

\paragraph{Consent and Platform Compliance.}
Reddit’s Terms of Service\footnote{\url{https://redditinc.com/policies}} permit the use of publicly available content for research purposes under specified conditions, and our data collection and redistribution practices comply with these terms. Our approach is consistent with established academic precedent for Reddit-based datasets and large-scale public text corpora \citep{baumgartner2020pushshift,bhargav_tot_2023,shen2023modeling}.
While obtaining explicit consent from individual users would be even better, this is infeasible in practice due to the pseudonymous nature of accounts and the time elapsed since posting. As in prior work, we therefore rely on platform-level consent and widely accepted research norms and practices in computational social science and natural language processing.

\paragraph{Copyright and Music Content.}

MusicSem does not directly redistribute any audio content. Instead, it provides references to musical works via Spotify identifiers, song titles, and artist names extracted from user posts. This design respects copyright constraints while allowing researchers to locate audio through legally authorized platforms. Users are responsible for ensuring compliance with applicable copyright laws when accessing audio.

We recognize that long-term availability of referenced music remains a broader challenge in music research. Compared to datasets that rely on direct hosted URLs (for example, MusicCaps \citep{agostinelli_musiclm_2023} only links to YouTube URLs that may be removed over time), the use of platform-agnostic identifiers improves resilience to link decay and copyright takedowns, although permanence cannot be fully guaranteed. 
\section{Evaluating Multimodal Music Models Using MusicSem}
\label{sec:evaluation}

We now conduct a comprehensive set of experiments to evaluate a wide range of cross-modal music retrieval, text-to-music generation, and music-to-text generation models using MusicSem. These experiments provide concrete examples of tasks for which MusicSem can be used, while also demonstrating the value of this dataset as a benchmark for semantics-aware multimodal music model evaluation. We first present experiments on cross-modal music retrieval in Section~\ref{sec:exp-retrieval}, followed by music-to-text and text-to-music generation experiments in Sections~\ref{sec:exp-music-to-text} and~\ref{sec:exp-text-to-music}, respectively. Finally, Section~\ref{sec:finetuning} goes beyond evaluation by reporting preliminary fine-tuning experiments that illustrate the potential of MusicSem for model adaptation. For reproducibility, we publicly release our source code on GitHub at: \url{https://github.com/Rsalganik1123/MusicSem}.

\subsection{Cross--Modal Music Retrieval}
\label{sec:exp-retrieval}

\paragraph{Setting.}
We first benchmark representative models on a \textit{cross-modal music retrieval task}, a core application of multimodal music representation learning \citep{wu_largescale_2023,wu_clamp3_2025}. Specifically, we focus on the text-to-audio retrieval setting, where each query consists of a textual description from a language--audio pair, and the model observes only the text modality. The goal is to retrieve the corresponding audio track from a pool of candidate audio samples in the dataset. To perform retrieval, models embed both text queries and audio candidates into a shared embedding space and rank the audio tracks according to their similarity to the query embedding vector. A retrieval is deemed correct if the audio track originally paired with the query text is ranked among the top results \citep{wu_largescale_2023}.

We evaluate four competitive models for cross-modal retrieval from the recent literature: CLAP~\citep{wu_largescale_2023}, ImageBind~\citep{girdhar_imagebine_2023}, LARP~\citep{salganik_larp_2024}, and ClaMP3~\citep{wu_clamp3_2025}, along with a random retrieval baseline. For each model, we report standard retrieval metrics \citep{carterette2011overview}, including Mean Reciprocal Rank (MRR), Recall@$K$, and Normalized Discounted Cumulative Gain@$K$ (NDCG@$K$), with $K \in \{1, 5, 10\}$ denoting the number of top-ranked candidate audio tracks returned by the model.
We compute these metrics on MusicSem as well as on two other widely used language–audio music datasets, MusicCaps~\citep{agostinelli_musiclm_2023} and Song Describer~\citep{manco_describer_2023}, to enable comparative analysis across datasets.

For clarity and brevity, Appendix~\ref{app:experimentalsettings} provides detailed descriptions of all evaluated methods, links to their implementations, information on dataset splits and hyperparameters, and the computational resources employed.

\paragraph{Insight 1.1: Different datasets induce different model rankings.}
The results in Table~\ref{table:result_text_to_audio_ret} show that the relative performance of models varies substantially across datasets. On MusicCaps, CLAP achieves the strongest performance (for example, with a top 22.60\% Recall@10 score), whereas on both Song Describer and MusicSem, ClaMP3 emerges as the best-performing model (for example, with a top 26.84\% Recall@10 score on MusicSem).

We attribute these differences primarily to mismatches between the training data of the models and the characteristics of the evaluation datasets. CLAP was originally trained on a mixture of music and general audio content, including ambient sounds, whereas ClaMP3 is designed specifically for music representation learning. This distinction aligns with the nature of the evaluated datasets. MusicCaps contains audio clips sourced from YouTube and overlaps with AudioSet~\citep{audioset}, which includes a wide range of non-musical and ambient audio. In contrast, both Song Describer and MusicSem rely exclusively on studio-quality music recordings that are largely free of ambient noise.

The observed variation in model rankings across datasets highlights limitations in the generalization capabilities of current multimodal music understanding models and suggests substantial room for improvement in developing representations that transfer robustly across diverse audio domains.

\begin{table} [t] %\scriptsize
\centering
\caption{Evaluation results on the text-to-music retrieval task. Best performance for each metric within a dataset is shown in bold, and second-best results are underlined. All scores are reported as percentages. In this retrieval setting, NDCG@1 is equivalent to Recall@1 and is therefore not reported in the table.}
\resizebox{\linewidth}{!}{
\begin{tabular}{c|c| cccccc  }
\toprule 
\textbf{Dataset} &
\textbf{Model} & $\mathbf{Recall@1 \uparrow}$ & $\mathbf{Recall@5 \uparrow}$ & $\mathbf{Recall@10 \uparrow}$ & $\mathbf{NDCG@5 \uparrow}$ & $\mathbf{NDCG@10 \uparrow}$ & $\mathbf{MRR \uparrow}$ \\
\midrule 
\multirow{5}{*}{\textbf{MusicCaps}} 
& Random & 0.04 & 0.18 & 0.36 & 0.10 & 0.16 & 0.31 \\ 
& LARP & 0.14 & 0.49 & 0.98 & 0.30 & 0.45 & 0.62 \\ 
& CLAP & \textbf{5.84} & \textbf{15.57} & \textbf{22.60} & \textbf{10.73} & \textbf{12.99} & \textbf{11.60} \\ 
& ImageBind & \underline{3.15} & \underline{10.18} & \underline{14.91} & \underline{6.72} & \underline{8.25} & \underline{7.23}  \\
& CLaMP3 & 2.73 & 8.82 & 13.65 & 5.81 & 7.32 & 9.07 \\ 
\midrule
\multirow{5}{*}{\textbf{Song Describer}}
& Random & 0.14 & 0.71 & 1.41 & 0.41 & 0.64 & 1.01 \\ 
& LARP & 0.36 & 1.72 & 2.62 & 1.05 & 1.29 & 1.61 \\ 
& CLAP & \underline{4.61} & \underline{17.3} & \underline{27.67} & \underline{11.20} & \underline{14.54} & \underline{12.41} \\ 
& ImageBind & 4.43 & 13.02 & 20.71 & 8.72 & 11.16 & 9.84 \\ 
& CLaMP3 & \textbf{10.49} & \textbf{27.31} & \textbf{38.61} & \textbf{19.21} & \textbf{22.84} & \textbf{19.83} \\ 
\midrule
\multirow{5}{*}{\textbf{MusicSem}}
& Random & 0.21 & 1.05 & 2.11 & 0.62 & 0.96 & 1.42 \\ 
& LARP & 0.22 & 1.02 & 3.07 & 0.54 & 1.22 & 1.47 \\ 
& CLAP & 0.82 & 5.74 & 9.84 & 3.54 & 4.74 & 4.65 \\ 
& ImageBind & \underline{2.05} & \underline{5.94} & \underline{11.07} & \underline{3.83} & \underline{5.48} & \underline{5.24}  \\ 
& CLaMP3 & \textbf{7.79} & \textbf{18.85} & \textbf{26.84} & \textbf{13.65} & \textbf{16.21} & \textbf{14.68} \\ 
\bottomrule
\end{tabular}
}
\label{table:result_text_to_audio_ret}
\end{table}

\paragraph{Insight 1.2: MusicSem is more challenging than existing datasets.}
Comparing performance on MusicSem and Song Describer, we observe that nearly all evaluated models perform worse on MusicSem, despite its smaller candidate set (480 test audio tracks) compared to Song Describer (note that in retrieval tasks, larger candidate sets typically lead to lower performance \citep{carterette2011overview}). 
This indicates that MusicSem constitutes a more challenging benchmark for multimodal models.

A similar trend appears when comparing MusicSem with MusicCaps. Although MusicCaps involves a larger candidate set, model performance on MusicCaps is only slightly lower than on Song Describer, and in some cases higher than on MusicSem. This suggests that the increased difficulty of MusicSem cannot be attributed solely to candidate set size.

Overall, these results indicate that MusicSem introduces additional semantic challenges for cross-modal retrieval, likely due to its richer and more nuanced textual descriptions. This finding suggests that current multimodal music representation models remain limited in their ability to capture fine-grained musical semantics, and that MusicSem provides a valuable benchmark for studying and advancing semantic understanding in cross-modal music retrieval in future research.

\subsection{Music-to-Text Generation}
\label{sec:exp-music-to-text}

\paragraph{Setting.}
MusicSem is also well suited for evaluating cross-modal generation tasks, including \textit{music-to-text generation}, also referred to as music captioning \citep{manco_describer_2023, liu_music_2024, wu_futga_2024}. In this task, the model is provided with an audio recording of a musical work as input and is required to generate a natural language description that captures its semantic content, including descriptive and contextual aspects of the music.

We evaluate three competitive music captioning models from the recent literature: MU-LLaMA~\citep{liu_music_2024}, LP-MusicCaps~\citep{doh_lpmusiccaps_2023}, and FUTGA~\citep{wu_futga_2024}. As for cross-modal music retrieval, experiments are conducted on MusicCaps, Song Describer, and our proposed dataset, MusicSem. We report commonly used  evaluation metrics from natural language processing, including BLEU (B)~\citep{papineni_bleu_2002}, METEOR (M)~\citep{banerjee_meteor_2005}, ROUGE (R)~\citep{lin_rouge_2004}, CIDEr~\citep{vedantam_cider_2015}, and BERTScore (BERT-S)~\citep{bert}, which are standard for evaluating music captioning models.

For clarity and brevity, Appendix~\ref{app:experimentalsettings} provides detailed descriptions of all evaluated methods, links to their implementations, information on dataset splits and hyperparameters, and the computational resources employed. The appendix also includes a discussion of the evaluation metrics and the intuitions underlying them.

\paragraph{Insight 2.1: Model performance varies across datasets and metrics.}

The results in Table~\ref{tab:m2t} show that model performance for music-to-text generation varies substantially across datasets. On both MusicCaps and MusicSem, LP-MusicCaps achieves the strongest overall performance according to most of the reported  metrics (for example, with a top 53.21 $\text{BLEU}_1$ score on MusicCaps), whereas on Song Describer, MU-LLaMA tends to outperform the other models. This variability is consistent with the performance inconsistencies observed in the cross-modal retrieval task and further suggests that existing music captioning models exhibit limited generalization across datasets. Improving robustness and generalization therefore remains a key challenge for music-to-text generation.

\begin{table}[t] %\scriptsize
\centering 
\caption{Evaluation results on the music-to-text generation task. Best performance for each
metric within a dataset is shown in bold.} \label{tab:m2t}
\resizebox{\linewidth}{!}{
\begin{tabular}{c|c| ccccccc c c}
\toprule 

\textbf{Dataset} &
\textbf{Model} &$\mathbf B_1 \uparrow$ & $\mathbf B_2 \uparrow$ & $\mathbf B_3 \uparrow$ & $\mathbf M$ ↑& $\mathbf R \uparrow$ & \textbf{CIDEr} $\mathbf \uparrow$ & \textbf{BERT-S} $\mathbf \uparrow$ \\
\midrule 
\multirow{3}{*}{\textbf{MusicCaps}} & 
LP-MusicCaps & \textbf{53.21} & \textbf{47.28} &  \textbf{44.60} & \textbf{51.90} & 3.35 & \textbf{384.72} & \textbf{90.47} \\ 
& MU-LLaMA & 1.35 & 0.55 & 0.22 & 40.22 & 11.27 & 0.09 & 80.47 \\ 
% & MusiLingo & x & x & x & x & x & x & x \\ 
& FUTGA& 8.80 & 3.07 & 1.19 & 44.77 & \textbf{11.90} & 2.63e-17 & 81.67 \\ 
\midrule
\multirow{3}{*}{\textbf{Song Describer}} &
LP-MusicCaps & 9.51 & 3.07 & 0.94 & \textbf{8.90} & 10.45 & 1.03 & \textbf{84.40} \\ 
& MU-LLaMA & \textbf{12.03} & \textbf{4.73} &  \textbf{1.73} & 8.72 & \textbf{13.00} & \textbf{3.59} & 83.51 \\ 
% & MusiLingo& x & x & x & x & x & x & x \\ 
& FUTGA& 3.39 & 1.28 & 0.43 & 8.72 & 6.30 & 3.58e-30 & 82.55 \\ 
\midrule
\multirow{3}{*}{\textbf{MusicSem}} 
& LP-MusicCaps & \textbf{11.57} & \textbf{3.05} & \textbf{0.72} & \textbf{20.59} & 9.54 & 0.77 & \textbf{82.13} \\ 
& MU-LLaMA & 4.11 & 1.41 & 0.51 & 22.33 & \textbf{10.57} & \textbf{0.92} & 81.63 \\ 
% & MusiLingo& x & x & x & x & x & x & x \\ 
& FUTGA& 4.82 & 1.50 & 0.44 & 22.23 & 7.48 & 0.01 & 80.93 \\ 
\bottomrule
\end{tabular}
}
\end{table}

\begin{table}
\caption{Semantic analysis of music-to-text generation on MusicSem. We report the proportion of captions containing at least one element from each semantic category, for both ground-truth annotations in the MusicSem test set and model-generated captions.}
\label{tab:m2t_sem}
\centering
\resizebox{\linewidth}{!}{
\begin{tabular}{c|ccc|c}
    \toprule
    \textbf{Category of Semantics} & \textbf{LP-MusicCaps} & \textbf{MU-LLaMA} & \textbf{FUTGA} & \textbf{Ground Truth in MusicSem}  \\ 
    \midrule
    Descriptive & 100\% & 99\% & 100\% & 83\%  \\ 
    Contextual & 2\% & 1\% & 0\% & 17\%  \\ 
    Situational & 42\% & 0\% & 1\% & 38\% \\
    Atmospheric & 78\% & 3\% & 91\% & 62\% \\
    Metadata & 32\% & 2\% & 34\% & 15\% \\ 
    \bottomrule
\end{tabular}
}
\end{table}

\paragraph{Insight 2.2: Performance inconsistencies are driven by semantic diversity across datasets.}
To further investigate the observed performance inconsistencies, we analyze the distribution of semantic categories present in both the ground-truth annotations of the MusicSem test set and the captions generated by each model, as reported in Table~\ref{tab:m2t_sem}. This analysis shows that the relatively strong performance of LP-MusicCaps correlates with its higher coverage of atmospheric, situational, and contextual semantics in its generated captions. Among the evaluated models, LP-MusicCaps exhibits the highest proportion of these semantic categories in its output.

At the same time, we observe that all models are heavily skewed toward producing descriptive captions, while only a small fraction of generated text captures contextual, situational, or atmospheric elements that are prominent in Reddit-based annotations. This imbalance highlights the difficulty of generating rich and faithful semantic descriptions of music using current state-of-the-art models. These results suggest that MusicSem, with its emphasis on diverse and context-rich semantics, provides a valuable benchmark for identifying and addressing these limitations in future work.

\paragraph{Insight 2.3: Higher scores do not always imply semantic correctness.}

\begin{figure}[t]
    \centering
    \includegraphics[width=\linewidth]{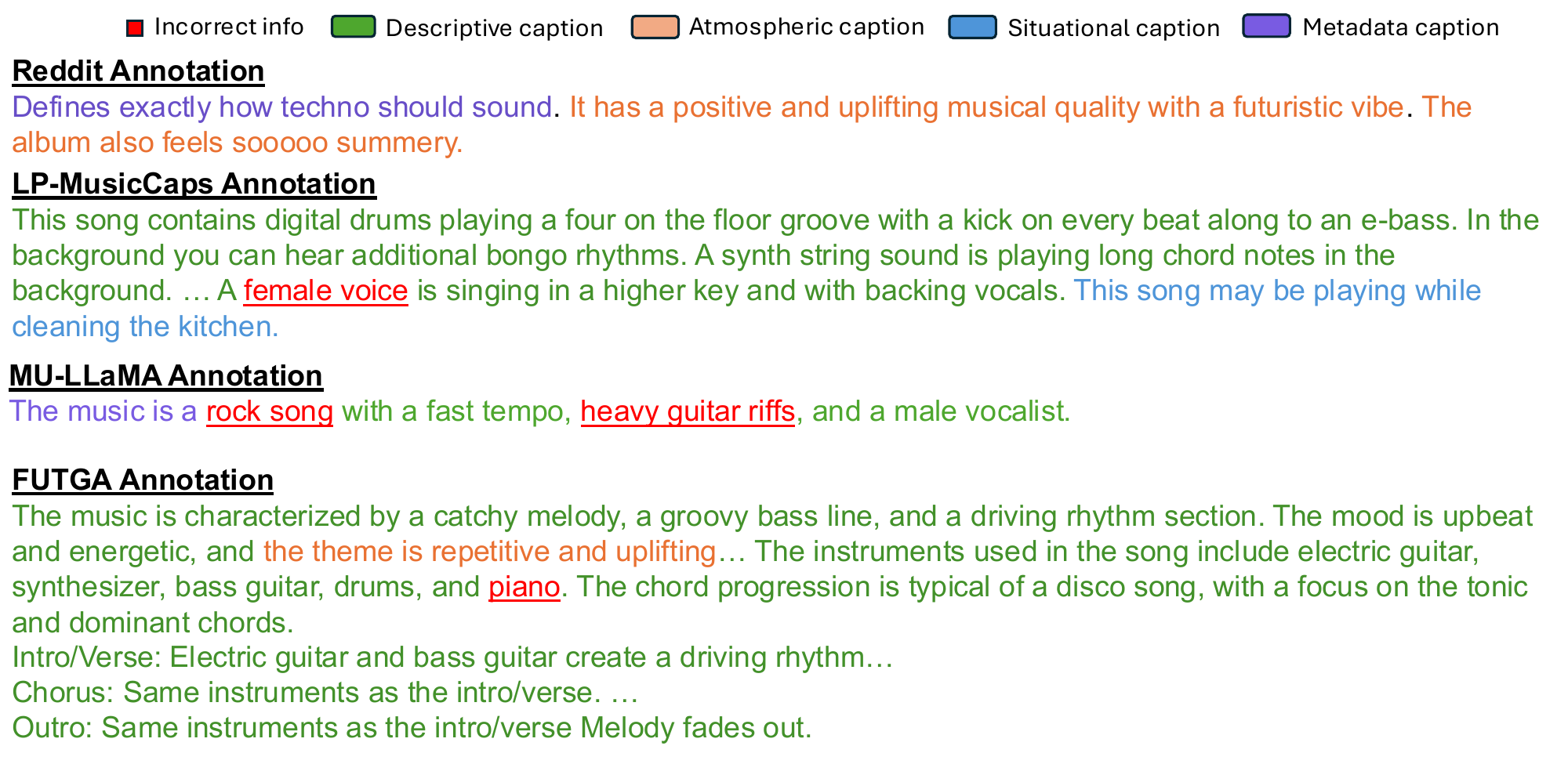}
\caption{Case study of music-to-text generation evaluation using MusicSem. The reference song is \emph{While Others Cry} by The Future Sound of London. All models produce captions containing factual inaccuracies and primarily focus on descriptive attributes.}
    \label{fig:case_study_m2t}
\end{figure}

Interestingly, a close inspection of model outputs reveals some limitations of metric-based evaluation. Although LP-MusicCaps attains the highest scores, qualitative analysis challenges this conclusion. Figure~\ref{fig:case_study_m2t} presents a representative case study on MusicSem, comparing the ground-truth annotation with captions generated by each model. In this example, FUTGA produces a more detailed and semantically accurate description of the audio, yet receives lower objective scores due to reduced $n$-gram overlap caused by longer and more expressive outputs. In contrast, MU-LLaMA generates shorter captions that are largely incorrect, but nevertheless achieves scores comparable to FUTGA, likely due to superficial lexical overlap.

Moreover, despite seemingly strong quantitative performance across models, each generated caption in the case study contains at least one factually incorrect description of the input music. This highlights a persistent gap between objective evaluation metrics and true semantic understanding, and indicates that current state-of-the-art models remain limited in their ability to capture fine-grained and faithful musical semantics.

\subsection{Text-to-Music Generation}
\label{sec:exp-text-to-music}

\paragraph{Setting.}
The \textit{text-to-music generation} task consists in generating musical audio from a textual description. In this work, we focus on one of the most challenging settings, namely multi-track music generation, where the generated audio contains multiple instruments \citep{agostinelli_musiclm_2023,huang_noise2music_2023,schneider_mousai_2024,melechovsky_mustango_2024,lam_efficient_2023,copet_simple_2023,liu_audioldm2_2024}. We consider a one-shot prompting scenario, in which a single textual input is provided to the model without iterative refinement. We leave multi-turn interactive music generation \citep{lin_arrange_2024,ronchini_paguri_2024,zhang_instruct_2024} for future work.

We evaluate six competitive text-to-music generation models: MusicLM \citep{agostinelli_musiclm_2023}, Stable Audio \citep{evans_fast_2024}, MusicGen \citep{copet_simple_2023}, AudioLDM2 \citep{liu_audioldm2_2024}, Mustango \citep{melechovsky_mustango_2024}, and the proprietary Mureka generative tool\footnote{\url{https://www.mureka.ai/}}. For each model, we report several metrics, grouped into three complementary dimensions: (i) audio quality, measured by the Fréchet Audio Distance (FAD) \citep{kilgour_frechet_2019,gui_adapting_2024}; (ii) audio diversity, measured by the Kullback--Leibler Divergence (KLD) \citep{kld} and Vendi Score (VS) \citep{friedman_vendi_2023}; and (iii) text--audio fidelity, measured by the CLAP score (CS) \citep{wu_largescale_2023}. Again, we conduct experiments on MusicCaps, Song Describer, and our proposed dataset, MusicSem.

For clarity and brevity, Appendix~\ref{app:experimentalsettings} provides detailed descriptions of all evaluated methods, links to their implementations, information on dataset splits and hyperparameters, and the computational resources employed. The appendix also includes a discussion of the evaluation metrics and the intuitions underlying them.

\paragraph{Insight 3.1: Each evaluation metric captures a different aspect of generation quality.}
First, different variants of the Fréchet Audio Distance (FAD) yield substantially different rankings among models. Given a reference embedding model (indicated by the subscript, where V, M, and E correspond to VGG \citep{simonyan_very_2015}, MERT \citep{li_mert_2024}, and Encodec \citep{defossez_high_2022}, respectively) and a reference dataset (indicated by the superscript, where MC and FMA refer to MusicCaps \citep{agostinelli_musiclm_2023} and the Free Music Archive (FMA) \citep{fma}, respectively), FAD measures the discrepancy between the distributions of real and generated audio embeddings. Consistent with the observations of~\citet{gui_adapting_2024}, we find that FAD values computed using different reference models can differ by orders of magnitude, leading to markedly different relative rankings across competing systems.

Although the proprietary model Mureka achieves the strongest overall FAD scores, the identity of the second-best model varies considerably depending on the choice of reference model and dataset. This variability suggests that non-proprietary models still exhibit substantial limitations in generating consistently high-quality music. Second, models that achieve low (better) FAD scores do not necessarily obtain high Vendi Scores, indicating a trade-off between audio quality and diversity. Achieving both high fidelity and high diversity remains a challenging open problem in text-to-music generation.

Finally, we observe discrepancies between performance measured by  objective metrics (Table~\ref{table:t2m}) and semantic sensitivity results (Table~\ref{tab:sensitivity_t2m}). For instance, Stable Audio performs strongly on semantic sensitivity tests but scores poorly on conventional metrics, whereas Mustango exhibits the opposite trend. These inconsistencies highlight the difficulty of jointly optimizing semantic alignment and traditional audio-based evaluation criteria.

Overall, these findings underscore the complexity of evaluating text-to-music generation and demonstrate that no single metric provides a complete assessment of model performance. The inclusion of semantic sensitivity as an evaluation dimension introduces additional challenges for current methods, which are made explicit with MusicSem.

\begin{table}[t] 
\centering
\caption{Evaluation results on the text-to-music generative task. Best performance for each
metric within a dataset is shown in bold, and second-best results are underlined. Note: Experiments for Mureka on MusicCaps could not be completed due to unresolved issues (at the time of writing) with the Mureka API, which prevent automated access to audio data from their website. For completeness, we nonetheless report the results obtained for Mureka on MusicSem and Song Describer.}
\label{table:t2m}
\resizebox{\linewidth}{!}{\begin{tabular}{c|c|ccccccc}
\toprule
\textbf{Dataset} & \textbf{Model} & \textbf{FAD$_\text{V}^\text{MC}$} ↓ & \textbf{FAD$_\text{V}^\text{FMA}$} ↓ & \textbf{FAD$_\text{M}^\text{FMA}$} ↓ & \textbf{FAD$_\text{E}^\text{FMA}$} ↓ & \textbf{KLD} ↓ & \textbf{Vendi} ↑ & \textbf{CS} ↑ \\
\midrule

\multirow{6}{*}{\textbf{MusicCaps}} 
& MusicLM      & 5.70 & 21.57 & 87.39 & 249.72 & 1.79 & \underline{1.55} & 0.28 \\
& Stable Audio & 6.97 & \textbf{15.60} & 82.21 & 377.02 & 1.90 & 1.31 & \underline{0.31} \\
& MusicGen     & 7.03 & \underline{16.29} & 73.22 & 354.07 & \underline{0.90} & \textbf{1.57} & 0.29 \\
& AudioLDM2    & \underline{3.29} & 19.31 & \underline{60.02} & \underline{202.11} & \textbf{0.61} & \textbf{1.57} & \textbf{0.36} \\
& Mustango     & \textbf{1.27} & 22.96 & \textbf{55.84} & \textbf{161.47} & 1.51 & 1.48 & 0.27 \\
& Mureka       & - & - & - & - & - & - & - \\
\midrule

\multirow{6}{*}{\textbf{Song Describer}} 
& MusicLM      & 7.20 & 20.59 & 87.12 & 241.95 & 0.89 & 1.49 & 0.28 \\
& Stable Audio & 4.42 & 14.90 & 79.16 & 341.92 & 1.07 & 1.29 & 0.31 \\
& MusicGen     & 2.64 & \underline{14.60} & 65.74 & 354.07 & \underline{0.66} & \textbf{1.50} & \textbf{0.35} \\
& AudioLDM2    & 2.74 & 17.19 & 57.88 & 184.03 & \textbf{0.62} & \underline{1.48} & \underline{0.34} \\
& Mustango     & 2.58& 18.50 & \underline{56.69} & \underline{170.27} & 1.48 & 1.46 & 0.29 \\
& Mureka       & \textbf{2.42} & \textbf{9.85} & \textbf{35.58} & \textbf{47.84} & 1.38 & 1.38 & 0.23 \\
\midrule

\multirow{6}{*}{\textbf{MusicSem (Ours)}} 
& MusicLM      & 7.25 & 22.57 & 86.97 & 248.42 & 1.00 & \underline{1.46} & 0.27 \\
& Stable Audio & 5.50 & 14.96 & 79.35 & 342.53 & 1.15 & 1.28 & \textbf{0.31} \\
& MusicGen     & 3.75 & \underline{14.67} & 68.11 & 229.29 & \underline{0.74} & \textbf{1.50} & \underline{0.30} \\
& AudioLDM2    & \underline{3.47} & 17.66 & 57.71 & 181.11 & \textbf{0.55} & 1.46 & 0.28 \\
& Mustango     & 5.06 & 19.15 & \underline{55.11} & \underline{157.32} & 1.46 & 1.41 & 0.20 \\
& Mureka       & \textbf{2.70} & \textbf{9.69} & \textbf{34.75} & \textbf{44.75} & 1.40 & 1.33 & 0.18 \\

\bottomrule
\end{tabular}}
\end{table}

\paragraph{Insight 3.2: Limitations of the CLAP score.}
The CLAP score is a widely used metric for evaluating the alignment between a textual prompt and its associated generated audio. However, our results reveal notable limitations of this metric. Specifically, we observe minimal performance differences across models when evaluated on canonical benchmark datasets and on MusicSem. This outcome is unexpected, as MusicSem contains substantially fewer descriptive annotations and richer contextual semantics, which would intuitively be reflected in differences in CLAP scores.

To further investigate this behavior, we leverage the semantic sensitivity metric defined in Equation~\eqref{eq:g} and compute cosine similarities between text embeddings produced by the CLAP text encoder. This analysis allows us to directly assess CLAP's sensitivity to semantic variations in textual prompts. As shown in Table~\ref{tab:clap_sen}, CLAP exhibits similarly low semantic sensitivity, suggesting a limited capacity to distinguish fine-grained semantic differences in text. Taken together, these findings indicate that the CLAP score is insufficient for capturing the rich and nuanced semantics present in MusicSem, and highlight the need for alternative or complementary evaluation metrics when assessing language--audio alignment.

\begin{table}
\caption{Sensitivity of the CLAP score on MusicSem. The superscripts $^d$, $^a$, $^s$, $^c$, and $^m$ refer to descriptive, atmospheric, situational, contextual, and metadata, respectively.}
\centering
\resizebox{0.45\linewidth}{!}{
\begin{tabular}{c|cc}
    \toprule
    \textbf{Category of Semantics} & \textbf{Metric} & \textbf{Score} \\
    \midrule
    Descriptive & $G^d$ & 0.55 \\
    Atmospheric & $G^a$ & 0.36 \\
    Situational & $G^s$ & 0.32 \\
    Contextual & $G^c$ & 0.29 \\
    Metadata & $G^m$ & 0.36 \\
    \bottomrule
\end{tabular}
}
\label{tab:clap_sen}
\end{table}

\section{Towards Fine-Tuning Multimodal Models using MusicSem}
\label{sec:finetuning}
\subsection{Usage Guidelines}\label{sec:usage_guidelines}
The primary goal of MusicSem is to support the development of multimodal music models with stronger semantic awareness; that is, models capable of capturing richer and more nuanced relationships between textual descriptions and audio content. While designing architectures that achieve effective semantic grounding remains an open research challenge and is beyond the scope of this paper, MusicSem was constructed with such use cases in mind. We therefore outline practical guidelines intended to serve as a road map for future practitioners.

MusicSem consists of language--audio pairs derived from user discussions, where each description reflects semantically meaningful attributes, impressions, or contextual information about a track. We provide two versions of the dataset: (1) a full version in which each user’s discussion constitutes a separate data point, and (2) a de-duplicated version in which each song appears only once. Because multiple users may provide conflicting or subjective characterizations of the same song, the full dataset introduces additional variability that may act as noise during training. Unless modeling such disagreement is explicitly desired, we recommend beginning with the de-duplicated version. Practitioners can use the \texttt{unique\_id} field to select a single instance per song (see Appendix~\ref{sec:datasetconstruction} for details on dataset fields). As no objective criterion exists for choosing a single ``best'' description, practitioners should consider how their selection strategy may influence the learned representations and tailor it to their modeling goals. After selecting a single textual description per song, MusicSem can be used in a manner similar to prior multimodal datasets such as MusicCaps~\cite{agostinelli_musiclm_2023} or Song Describer~\cite{manco_describer_2023} for contrastive pretraining or fine-tuning. In the following section, we present preliminary results demonstrating the use of MusicSem to fine-tune a CLAP-based model~\cite{wu_largescale_2023}.

\subsection{Preliminary Results}
The experiments presented so far aim to illustrate the value of MusicSem as a benchmark for semantically-aware evaluation of multimodal music models. As an opening step, we move beyond evaluation and report preliminary fine-tuning experiments that highlight the potential of MusicSem for enriching pre-trained models with semantic sensitivity.

We note that a comprehensive investigation of fine-tuning multimodal language--audio models could in itself constitute a full study, requiring careful consideration of training objectives and losses, semantic balancing during optimization, regularization to prevent catastrophic forgetting, and scaling effects \citep{church2021emerging,han2024parameter}. For these reasons, a full fine-tuning investigation using MusicSem is beyond the scope of this paper, our objective here is to show that MusicSem can enrich a pretrained multimodal model with fine-grained musical semantics, including contextual, situational, and atmospheric. 
\begin{table}[]
\centering
\caption{Effects of fine-tuning CLAP~\citep{wu_largescale_2023} on the MusicSem training set, evaluated in terms of cross-modal text-to-music retrieval performance and semantic sensitivity on the MusicSem test set. Best performance for each metric is shown in bold. Improvements are reported relative to the non--fine-tuned baseline. In this retrieval setting, NDCG@1 is equivalent to Recall@1 and is therefore not reported in the table.}
\label{tab:finetuning}
\resizebox{\linewidth}{!}{
\begin{tabular}{l|cccccc}
\toprule
\textbf{Retrieval Performance} 
& Recall@1 
& Recall@5 
& Recall@10 
& NDCG@5 
& NDCG@10 
& MRR \\
\midrule
No Fine-Tuning 
& 0.82 & 5.74 & 9.84 & 3.54 & 4.74 & 4.65 \\
Fine-Tuning on MusicSem 
& \textbf{4.51} & \textbf{11.48} & \textbf{17.42} & \textbf{8.06} & \textbf{9.95} & \textbf{9.16} \\
\textit{Relative Improvement}
& \textit{+450.00\%} & \textit{+100.00\%} & \textit{+77.03\%} & \textit{+127.68\%} & \textit{+109.92\%} & \textit{+96.77\%} \\
\midrule
\midrule
\textbf{Semantic Sensitivity} 
& $G^d$ 
& $G^a$ 
& $G^s$ 
& $G^c$ 
& $G^m$ & \\
\midrule
No Fine-Tuning 
& 0.55 & 0.36 & 0.32 & 0.29 & 0.36 & \\
Fine-Tuning on MusicSem 
& \textbf{0.64} & \textbf{0.38} & \textbf{0.41} & \textbf{0.43} & \textbf{0.38} & \\
\textit{Relative Improvement} 
& \textit{+16.36\%} &\textit{+5.56\%} & \textit{+28.13\%} & \textit{+48.28\%} & \textit{+5.56\%} & \\
\bottomrule
\end{tabular}
}
\vspace{-1em}
\end{table}

\paragraph{Setting.} We fine-tune CLAP~\cite{wu_largescale_2023}, a popular cross-modal music retrieval on the MusicSem training set. In our evaluation, we take the best performing checkpoint available on GitHub\footnote{\url{https://huggingface.co/lukewys/laion_clap/blob/main/630k-audioset-fusion-best.pt}} and fine-tune the model using the de-duplicated version of the MusicSem dataset for up to 200 epochs. We then select the checkpoint achieving the highest average semantic sensitivity score (as defined in Equation~\eqref{eq:g}) on a held-out validation set, reached at epoch 110. The selected model is then evaluated on the MusicSem test set. We report the effects of this fine-tuning procedure using two sets of metrics: (1) the retrieval performance metrics from Section~\ref{sec:exp-retrieval} (MRR, Recall@$K$, and NDCG@$K$, with $K \in \{1, 5, 10\}$), and (2) the semantic sensitivity scores for each semantic category ($G^d$, $G^a$, $G^s$, $G^c$, and $G^m$) in Table~\ref{tab:finetuning}.

\paragraph{Insight 4.1: Enhancing semantic awareness via MusicSem fine-tuning is feasible.}
Table~\ref{tab:finetuning} summarizes the results. The first key insight from our experiments is that the scale and semantic diversity of MusicSem are sufficient to fine-tune pretrained multimodal music models and that such fine-tuning can substantially enhance the semantic awareness of the pretrained model. Fine-tuning consistently improves sensitivity across all semantic categories, with especially pronounced gains for contextual (+48.28\%) and situational (+28.13\%) semantics. These two categories are central to the design of MusicSem and are largely underrepresented in prior datasets. The strong improvements observed on these dimensions suggest that the model is able to leverage the dataset’s richer semantic structure to better align audio representations with higher-level, user-expressed intent. This indicates that MusicSem not only serves as a challenging evaluation benchmark, but also provides training signals that are complementary to those found in existing large-scale language--audio corpora.

\paragraph{Insight 4.2: Fine-tuning using MusicSem can also improve performance.}
Table~\ref{tab:finetuning} further shows that, beyond improving semantic sensitivity, fine-tuning CLAP on MusicSem also yields substantial gains in text-to-music retrieval performance on the MusicSem test set. Across all reported retrieval metrics, performance improves markedly after fine-tuning. For instance, we observe a +96.77\% relative increase in MRR for the fine-tuned model compared to the baseline CLAP model without fine-tuning. These results indicate that improved semantic alignment can translate into tangible benefits on a downstream retrieval task, rather than remaining confined to semantic sensitivity metrics alone.

%\vspace{-1em}
\subsection{Limitations of this Dataset}
A central challenge in constructing datasets that capture semantically rich musical discourse lies in reliably aligning textual descriptions with the specific songs they reference. In informal, user-generated discussions, references to songs, artists, and albums are often ambiguous, implicit, or interwoven within longer narratives, making precise one-to-one text–song alignment difficult. Achieving such alignment would require robust named entity recognition (NER) tailored to the music domain, which remains an open problem for both human annotators and automated systems~\cite{epure2023human}. For example, entities such as \emph{Boston} may refer to a band, an album, or a song title depending on context.

Even with the domain expertise of the professionally trained musicians on our research team we found that manual disambiguation frequently depended on prior familiarity with the referenced works. Scaling this process to large corpora would therefore be labor-intensive and impractical, while current large language model–based approaches do not yet provide sufficient reliability. Consequently, we do not attempt to implement a comprehensive NER and disambiguation pipeline in the present work. Instead, we acknowledge this as a limitation of MusicSem and view more accurate large-scale entity resolution as an important direction for future research. To support such future improvements, we release the complete raw text alongside our annotations, enabling the community to revisit, refine, and extend the extraction process as more reliable tools and methods become available. 

% Finally, we wish to stress that the current version of MusicSem intentionally retains posts that (1) discuss multiple songs within a single utterance, (2) reference several works by the same artist, or (3) contain only brief or indirect mentions. While this design introduces additional noise and subjectivity, it more faithfully reflects the structure of real-world musical discourse. We believe this diversity may benefit future applications, such as personalized or conversational music systems~\cite{}, that must by able to account for the variability in organic musical discourse.  

\subsection{Future Applications of MusicSem}\label{sec:implications}
MusicSem enables several promising downstream applications. As LLM-based conversational agents become increasingly integrated into how users engage with creative content, researchers on streaming platforms have begun exploring prompt-conditioned playlist recommendation~\cite{doh2025talkplay, chaganty_beyond_2023,pasqua_language_2025,spotify_prompted_playlists_support_2026,spotify_ai_playlist_beta_2025}. The rich and nuanced textual descriptions in our dataset are explicitly designed to support modeling of subtle contextual semantics—such as situational context or mood-based cues—thereby enabling more refined personalization. Moreover, the diversity and subjectivity inherent in user discourse make MusicSem particularly well suited for conversational or interactive music agents~\cite{gao_advances_2021}. Rather than relying on a single canonical description of each song, such systems can adapt recommendations dynamically to individual preferences, interpretations, and mindsets. Finally, as music representation learning increasingly move from text-only settings toward broader multimodal scenarios, MusicSem can support tasks that integrate audio with additional modalities. For example, the dataset can be leveraged for music–video retrieval or captioning tasks by aligning audio representations with associated visual media~\cite{li_deep_2021, korbar_2018}. To facilitate such extensions, we link each entry to persistent platform identifiers (e.g., Spotify), enabling reliable cross-modal integration. Together, these directions highlight the broader potential of MusicSem as a foundation for semantically grounded and context-aware music intelligence systems.

\section{Conclusion and Future Work}
\label{sec:conclusion}

In this work, we introduced MusicSem, a semantics-aware language–audio dataset designed to better reflect how people naturally describe and engage with music. Motivated by the observation that existing multimodal music datasets fail to capture the breadth and nuance of human musical discourse, MusicSem is constructed from organic music-related discussions on the social media platform Reddit and comprises 32,493 language–audio pairs. Compared to prior datasets, MusicSem captures a richer spectrum of musical semantics and explicitly structures them through a taxonomy of five categories: descriptive, atmospheric, situational, metadata-related, and contextual. Beyond dataset construction and analysis, we used MusicSem to benchmark a wide range of competitive models across cross-modal music retrieval, music-to-text generation, and text-to-music generation tasks. Together, our findings highlight the importance of semantics-aware evaluation and position MusicSem as a valuable resource for studying and advancing more human-aligned multimodal music representation learning. We also emphasized that MusicSem can serve not only as an evaluation benchmark, but also as a supervision signal for fine-tuning multimodal music representation models, enabling them to better capture nuanced and human-centered musical semantics.

MusicSem is publicly released under the MIT License and is available on Hugging Face. The complete source code for dataset construction and experiment reproduction is publicly available on GitHub, and we have created an accompanying website that hosts documentation and visualizations, and will include a private leaderboard and a held-out test set in future iterations to support standardized evaluation. Throughout the paper, we have taken ethical and legal considerations seriously, and explicitly outlined the safeguards adopted during dataset construction, including measures related to user privacy and anonymity, consent and platform compliance, and copyright considerations. We believe these practices are essential for responsible dataset release and align with established community standards.  We also highlighted important properties of MusicSem that users should keep in mind, including considerations related to cultural representativeness and music genre distribution. Rather than treating these aspects as fixed characteristics, we designed MusicSem as an extensible resource: the full data extraction pipeline is released, enabling practitioners to augment the dataset with additional sources or genres according to their research needs.

Looking forward, MusicSem opens multiple avenues for future work. First, we plan to further expand the scale and scope of the dataset by incorporating additional music-related discussions and communities. Second, MusicSem currently focuses on discourse about music rather than lyrics or symbolic musical representations; extending the dataset to include such content could further benefit music representation learning. Third, we aim to broaden benchmarking efforts using MusicSem, including evaluations for controllable music generation and text-guided music recommendation. In addition, a natural next step is to conduct more in-depth studies of retrieval and generative models fine-tuned on MusicSem, exploring training strategies, scaling effects, and transfer to downstream tasks. Finally, insights from our experiments underscore the need for more comprehensive and semantically grounded evaluation metrics for language--audio alignment. Overall, MusicSem aims to serve as a foundation for future research on models that better understand the nuanced, contextual, and human-centered language through which people engage with music.


\impact{This work contributes to the broader effort of developing AI systems that better align with how humans naturally express intent, context, and meaning. While recent advances in multimodal learning have significantly improved performance on specialized tasks in domains such as music, substantial gaps remain between these systems and more general, human-centered understanding. By introducing MusicSem, a dataset grounded in organic and nuanced musical discourse, this work takes a step toward bridging that gap and advancing multimodal representation learning beyond purely descriptive or surface-level semantics.

Within the music domain, MusicSem has several concrete positive impacts. First, it provides a principled and extensible benchmark for evaluating multimodal models across retrieval and generation tasks, emphasizing semantic sensitivity rather than narrow performance metrics alone. Second, the semantic taxonomy and sensitivity analyses introduced in this work offer new tools for auditing and diagnosing model behavior, helping researchers better understand where current systems succeed or fail in capturing contextual, situational, and atmospheric aspects of music. Third, by releasing both the dataset and the full data construction pipeline, MusicSem is designed to remain relevant as the field evolves, enabling researchers and practitioners to adapt, extend, and repurpose the resource for emerging tasks in music understanding. 

At the same time, we acknowledge potential negative and societal considerations associated with this line of research. Generative music technologies raise well-documented concerns regarding artist displacement, authorship, and the ethical use of creative works. While MusicSem does not directly address issues such as memorization, copyright infringement, or economic impacts on artists, we recognize that any contribution to multimodal generation research exists within this broader context. We therefore emphasize responsible use, transparency, and ethical safeguards throughout the dataset’s construction and release, including attention to user privacy, consent, and copyright compliance. 

Overall, we view MusicSem as a research-oriented resource intended to support more semantically grounded and human-aligned music understanding systems. We hope that by foregrounding nuanced musical discourse and ethical considerations, this work encourages future research that advances technical capabilities while remaining attentive to the social and cultural dimensions of music and creative expression.}

\section*{Disclaimer}
Portions of this work were previously presented as a late-breaking demonstration at ISMIR 2025 \citep{musicsem-ismir} and at the NeurIPS 2025 AI4Music workshop \citep{musicsem-neurips}. These presentations were non-archival and did not appear in formal proceedings.

% Acknowledgements and Disclosure of Funding should go at the end, before appendices and references

\vskip 0.2in
\bibliography{references}

\clearpage
\appendix

\section{Dataset Construction and Prompts}
\label{app:construction}

This Appendix~\ref{app:construction} presents additional details of the MusicSem dataset construction pipeline, with a particular focus on the extraction steps and the prompts employed.

\subsection{Pseudocode for Dataset Construction Pipeline}

We present pseudocode for the complete dataset construction pipeline in Algorithm~1. In Lines~2–3, we filter posts within each thread by removing content authored by moderators and posts containing fewer than 20 characters. In Line~4, we perform semantic extraction using a large language model (LLM) guided by a predefined prompt (see Appendix~\ref{app:extract_prompt}).

In Line~6, we query the Spotify API to retrieve a unique identifier for each song mentioned in a thread. In Line~7, we apply the first hallucination check to verify alignment between the retrieved audio and the extracted song–artist pairs. In Line~8, we download the corresponding audio files for each validated track. In Line~9, we generate summarized captions from the extracted semantic categories, following formats similar to those used in \textit{MusicCaps}~\citep{agostinelli_musiclm_2023} and Song Describer~\citep{manco_describer_2023}. 

Finally, in Line~10, we conduct a second hallucination check using a different model to ensure that the generated summaries remain faithful to the extracted semantic categories (see Appendix~\ref{app:hallucination_prompt}). Overall, this pipeline yields 32,493 language–audio pairs. A visual overview of the entire dataset construction process is provided in Figure~\ref{fig:extract} of the paper.

\begin{algorithm}[h]
\caption{MusicSem Dataset Construction Pipeline}\label{alg:full}
\hspace*{\algorithmicindent} \textbf{Input}: thread name $T$, language models $\mathcal{M}_1$, $\mathcal{M}_2$  \\
\hspace*{\algorithmicindent} \textbf{Output} caption set $C$
\begin{algorithmic}[1]
\Procedure{Dataset Generation}{$T, \mathcal{M}$}
\State posts = Load\_Entire\_Thread(T) 
\State filtered = Length\_and\_Mod\_Filter(posts)
\State sa\_pairs, caption\_extracts = $\mathcal{M}_1$(filtered)
\State descriptive, atmospheric, situational, contextual, metadata = caption\_extracts
\State song\_ids = Spotify\_Metadata(sa\_pairs)
\State sa\_pairs = Hallucination\_Check1(sa\_pairs,fltrd)
\State mp3s = Spotify\_Audio(song\_ids)
\State final\_summaries = Summarize(sa\_pairs,caption\_extracts, mp3s)
\State filtered\_captions = Hallucination\_Check2(caption\_extracts, final\_captions, $\mathcal{M}_2$)
\EndProcedure
\end{algorithmic}
\end{algorithm}

\subsection{Formalizing Semantic Categories}\label{ap:sem_cat}
We constructed the semantic categories using a qualitative coding process inspired by standard user-study methodologies. Specifically, we closely read the five threads described in Section~\ref{sec:datasetconstruction} and annotated them according to the broad themes expressed in the discussions. Through iterative qualitative analysis, we identified recurring patterns and progressively consolidated them into higher-level conceptual categories that capture the dominant modes of musical discourse.
We emphasize that this taxonomy is not intended to be exhaustive, as the ways in which people describe and interpret music continue to evolve. Rather, it reflects the range of themes observed in our data. Additionally, we intentionally excluded lyrical analysis, as we consider it a distinct and substantial research direction in its own right.

\subsection{Extraction Prompt}
\label{app:extract_prompt}

Below, we present the prompt used to extract semantic content from raw Reddit posts. Following the semantic category definitions in Table~\ref{tab:categories}, the prompt decomposes the text into elements corresponding to each of the five categories. We also provide an example extraction to illustrate the expected output.
\newpage

\begin{lstlisting}[language=Python]
% Feature Extraction

Task Description
You are tasked with analyzing Reddit posts about music and extracting structured information into specific categories. When given a Reddit post discussing music, identify and extract the following:
Categories to Extract
Song/Artist pairs
(using the names of artists and their songs with unfixed form) some examples:

'Shake it Off by Taylor Swift'
'Radiohead's Weird Fishes'
'Genesis - Yes'
'Maroon5 [She Will Be Loved]'

Descriptive (using musical attributes)
This includes detailed observations about:

Instrumentation: 'I love the high pass filter on the vocals in the chorus and the soft piano in the bridge'
Production techniques: 'The way they layered those harmonies in the second verse is incredible'
Song structure: 'That unexpected key change before the final chorus gives me goosebumps'
Sound qualities: 'The fuzzy lo-fi beats with that vinyl crackle in the background'
Technical elements: 'The 6/8 time signature makes it feel like its swaying'

Contextual (using other songs/artists)
This includes meaningful comparisons such as:

Direct comparisons: 'Sabrina Carpenter's Espresso is just a mix of old Ariana Grande and 2018 Dua Lipa'
Influences: 'You can tell they've been listening to a lot of Talking Heads'
Genre evolution: 'It's like 90s trip-hop got updated with modern trap elements'
Sound-alikes: 'If you like this, you should check out similar artists like...'
Musical lineage: 'They're carrying the torch that Prince lit in the 80s'

Situational (using an activity, setting, or environment)
This includes relatable scenarios like:

Life events: 'I listened to this song on the way to quitting my sh**ty corporate job'
Regular activities: 'This is my go-to album for late night coding sessions'
Specific locations: 'Hits different when you're driving through the mountains at sunset'
Social contexts: 'We always play this at our weekend gatherings and everyone vibes to it'
Seasonal connections: 'This has been my summer anthem for three years running'

Atmospheric (using emotions and descriptive adjectives)
This includes evocative descriptions such as:

Emotional impacts: 'This song makes me feel like a manic pixie dream girl in a bougie coffeeshop'
Visual imagery: 'Makes me picture myself in a coming-of-age indie movie, running in slow motion'
Mood descriptions: 'It has this melancholic yet hopeful quality that hits my soul'
Sensory experiences: 'The song feels like a warm embrace on a cold day'
Abstract feelings: 'Gives me this feeling of floating just above my problems'

Lyrical (focusing on words and meaning)
This includes thoughtful commentary on:

Storytelling: 'The lyrics tell such a vivid story of lost love that I feel like I've lived it'
Wordplay: 'The clever double entendres in the chorus make me appreciate it more each listen'
Messaging: 'The subtle political commentary woven throughout the verses really resonates'
Personal connection: 'These lyrics seem like they were written about my own life experiences'
Quotable lines: 'That line 'we're all just stardust waiting to return' lives rent-free in my head'

Metadata (using information found in labels or research)
This includes interesting facts like:

Technical info: 'The song is hip-hop from the year 2012 with a bpm of 100'
Creation context: 'They recorded this album in a cabin with no electricity using only acoustic instruments'
Chart performance: 'It's wild how this underplayed track has over 500 million streams'
Artist background: 'Knowing the guitarist was only 17 when they recorded this makes it more impressive'
Release details: 'This deluxe edition has three bonus tracks that are better than the singles'

Sentiment (whether the person feels good or bad about the song)
Output Format
Return your analysis as a structured JSON with these categories:
Copy{
  'pairs': [(song_1, artist_1), (song_2, artist_2), ...], 
  'Descriptive': [], 
  'Contextual': [], 
  'Situational': [], 
  'Atmospheric': [], 
  'Lyrical': [], 
  'Metadata': [],
  'Sentiment': []
}
\end{lstlisting}

\begin{lstlisting}[language=Python]
% Example

%Input:
'I like Plastic Love by Mariya Takeuchi because of the funky, jazzy, retro vibes. I listen to this music at 3am when Im lonely because it romanticizes my loneliness and makes it meaningful. It helps me to enjoy my own loneliness. It has very distinctive synthesizer sounds in the chorus and leading bass lines in the bridge. The vocals are chill and blended. Another song that sounds very similar is Once Upon a Night by Billyrrom or Warm on a Cold Night by Honne. The genre is like City Pop which describes an idealized version of a city.'

% Output:
Copy{
  'pairs': [('Plastic Love', 'Mariya Takeuchi'), ('Once Upon a Night', 'Billyrrom'), ('Warm on a Cold Night', 'HONNE')],
  'Situational': ['3am when Im lonely'],
  'Descriptive': ['funky', 'jazzy', 'retro vibes', 'distinctive synthesizer in chorus', 'leading bass lines in bridge', 'chill and blended vocals', 'genre of City Pop'],
  'Atmospheric': ['romantic loneliness', 'vulnerability', 'kind of sad in a good way', 'acting heartbroken', 'idealized version of a city'],
  'Contextual': ['Plastic Love sounds similar to Once Upon a Night', 'Plastic Love sounds similar to Warm on a Cold Night'],
  'Metadata': ['funky', 'jazzy', 'retro vibes', 'genre of City Pop']
}    
\end{lstlisting}

\subsection{Verification and Hallucination Check Prompt}
\label{app:hallucination_prompt}

Below, we present the prompt used to validate the outputs of the extraction and summarization stages. A secondary language model is employed to detect hallucinations by checking for inconsistencies between the extracted semantic tags and their sentence-level summarization. The model is provided with two illustrative examples: one negative example (i.e., containing no hallucinations) and one positive example (i.e., containing hallucinations). Our tests show that including both examples substantially improves the model’s ability to identify hallucinations.

\begin{lstlisting}[language=Python]
% Getting summarizations
# Summarization task

Write a sentence which combines the associated sentence fragments.
Please do not add anything other than the information given to you.

Your description should:
- Be maximum 4 sentences in length

Your description shouldn't:
- Add any additional information that is not present in the tags
- Include any information that is based on your own knowledge or assumptions

Example: 
  'Situational':  ['3am when Im lonely'],\
  'Descriptive':['funky', 'jazzy', 'retro vibes', 'distinctive synthesizer in chorus', 'leading bass lines in bridge', 'chill and blended vocals', 'genre of City Pop'],\
  'Atmospheric'': ['romantic loneliness', 'vulnerability', 'kind of sad in a good way', 'acting heartbroken', 'idealized version of a city'],\
  'Contextual': ['Plastic Love sounds similar to Once Upon a Night', 'Plastic Love sounds similar to Warm on a Cold Night'],\
  'Metadata': ['funky', 'jazzy', 'retro vibes', 'genre of City Pop']\

  Desired output: This song has funky, jazzy, retro vibes. I listen to this music at 3am when Im lonely because it romanticizes my loneliness and makes it meaningful. \
    It helps me to enjoy your own loneliness. It has very distinctive synthesizer sounds in the chorus and leading bass lines in the bridge. \
    The vocals are chill and blended.  The genre is like City Pop which describes an idealized version of a city.'  \

Tags:

{input_tags}

% Hallucination
# Hallucination Check Prompt for Generated Summary

## Instructions
Evaluate whether the generated summary contains hallucinations based on the provided features/tags from the original source. 
A hallucination is defined as information in the summary that is not present in or contradicts the features from the source material.

## Input Format
- **Original Features/Tags**: [List of key features/tags from the source material]
- **Generated Summary**: [The summary to be evaluated]

## Task
1. Compare each claim or statement in the summary against the original features/tags
2. Identify any information in the summary that:
   - Is not supported by the original features/tags
   - Contradicts the original features/tags
   - Represents an embellishment beyond what can be reasonably inferred
3. **The output should be in JSON format.**

## Output Format
```
{{
"hallucination_detected": [True/False],
}}
```

## Example 1
**Input Data**:
{{
  "original_features": {{
    'situational':  ['3am when Im lonely'],
    'descriptive':['funky', 'jazzy', 'retro vibes', 'distinctive synthesizer in chorus', 'leading bass lines in bridge', 'chill and blended vocals', 'genre of City Pop'],
    'atmospheric'': ['romantic loneliness', 'vulnerability', 'kind of sad in a good way', 'acting heartbroken', 'idealized version of a city'],
    'contextual': ['Plastic Love sounds similar to Once Upon a Night', 'Plastic Love sounds similar to Warm on a Cold Night'],
  }},
  "generated_summary": 'funky, jazzy, retro vibes. I listen to this music at 3am when Im lonely because it romanticizes my loneliness and makes it meaningful. 
    It helps me to enjoy your own loneliness. It has very distinctive synthesizer sounds in the chorus and leading bass lines in the bridge. 
    The vocals are chill and blended.  The genre is like City Pop which describes an idealized version of a city.'  
}}


**Expected Output**:
```
{{
"hallucination_detected": False,
}}

## Example 2
**Input Data**:
{{
  "original_features": {{
    'situational':  ['when I'm quitting my corporate job'],
    'descriptive':['angry punk guitar', 'killer drums', 'harcore vocal processing', 'distortion'],
    'atmospheric'': ['pumped up vibes', 'makes me want to take charge of my life'],
    'contextual': [''],
  }},
  "generated_summary": 'This song makes me happy. It has a soft and exciting vibe with killer drums. I listen to this song at parties or festivals when I feel positive.'
}}

**Expected Output**:
```
{{
"hallucination_detected": True,
}}
```

**Input Data**:
```
{{
  "original_features": {features},
  "generated_summary": {summary}
}}
```
**Expected Output**:
```
\end{lstlisting}

\section{Experimental Settings}
\label{app:experimentalsettings}

This Appendix~\ref{app:experimentalsettings} provides additional details on the experiments presented in the paper, including descriptions of all evaluated methods, the set of hyperparameters used, information on training and test splits, details about the computational resources employed, and a discussion of the evaluation metrics.

\subsection{Overview of Models}
\begin{table}[h]
\centering
\caption{Overview of all models evaluated in this work. \textit{Hier.}, \textit{Trans.}, \textit{Diff.}, and \textit{Co-List.} denote Hierarchical, Transformer, Diffusion, and Co-Listing, respectively.}
\label{tab:all_models}
\resizebox{\linewidth}{!}{
\begin{tabular}{clcccccc}
\toprule
\textbf{Task} & 
\textbf{Name} & \textbf{Date} & \textbf{Architecture} & \textbf{Text Conditioner} & \textbf{Length} & \textbf{Sample Rate} & \textbf{Proprietary} \\
\midrule
\multirow{6}{*}{\textbf{Text-to-Music}}& MusicLM \citep{agostinelli_musiclm_2023} & 2023 & Hier. Trans. + SoundStream & w2v-BERT \citep{chung_wav2bert_2021} & variable & 24kHz   &  \\
& AudioLDM 2 \citep{liu_audioldm2_2024} & 2023 & VAE + 2D U-Net & CLAP \citep{wu_largescale_2023} & variable & 16kHz   &  \\
& Stable Audio \citep{evans_fast_2024} & 2023 & VAE + 2D U-Net & CLAP \citep{wu_largescale_2023} & up to 95s & 48kHz &  \\
& MusicGen \citep{copet_simple_2023} & 2024 & AE + 1D U-Net & FLAN-T5 \citep{chung_scaling_2022} & 10s & 48kHz &  \\
& Mustango \citep{melechovsky_mustango_2024} & 2024 & VAE + 2D U-Net & FLAN-T5 \citep{chung_scaling_2022} & 10s & 16kHz   &  \\
& Mureka  & 2024&- &- & -&- & \checkmark \\
\midrule
\textbf{Task} & 
\textbf{Name} & \textbf{Year} & \textbf{Architecture} & \textbf{Audio Conditioner} & \textbf{Length} & \textbf{Sample Rate} & \textbf{Proprietary} \\
\midrule 
\multirow{3}{*}{\textbf{Music-to-Text}} 
& MU-LLaMA \citep{liu_music_2024} & 2024 & Diff. Trans. &MERT \citep{li_mert_2024} & 60s & 16kHz &  \\
&LP-MusicCaps \citep{liu_music_2024} & 2023 & Trans. & BART \citep{lewis_bart_2020} & 10s & 16kHz &  \\
&FUTGA \citep{wu_futga_2024} & 2024 & Hier. Trans. + VAE & Whisper\citep{radford2022robustspeechrecognitionlargescale} & 240s & 16kHz &  \\
\midrule
\textbf{Task} & 
\textbf{Name} & \textbf{Year} & \textbf{Architecture} & \textbf{Modalities} & \textbf{Length} & \textbf{Sample Rate} & \textbf{Proprietary} \\
\midrule
\multirow{4}{*}{\textbf{Retrieval}}
&CLAP \citep{wu_largescale_2023} & 2023 & Contrastive Learning  & Text + Waveform  & - & 48kHz &  \\
&LARP \citep{salganik_larp_2024} & 2024 & Contrastive Learning  & Text + Waveform + Co-List. Graph &  & 48kHz&  \\
&ImageBind \citep{girdhar_imagebine_2023} & 2023 &  Contrastive Learning & Text + Image & - & 16kHz &   \\
&CLaMP3 \citep{wu_clamp3_2025} & 2024 & Contrastive Learning & Text + Image + Waveform & - & 24kHz &  \\
\bottomrule
\end{tabular} 
}
\end{table}

\subsection{Cross--Modal Music Retrieval Models}
\label{App_CrossModal_Ret}

\paragraph{CLAP \citep{wu_largescale_2023}} learns joint embeddings between audio clips and text descriptions through Contrastive Language–Image Pretraining~\citep{radford2021learning}, trained on 630K audio–text pairs. For audio data, CLAP first represents signals using log-Mel spectrograms at a sampling rate of 44.1\,kHz, and then employs CNN14~\citep{kong2020panns} (80.8M parameters), which is pretrained on AudioSet with approximately 2M audio clips. For text data, CLAP uses BERT~\citep{devlin2019bert} (110M parameters) to encode text descriptions, taking the [CLS] token embedding as the text representation. Both audio and text embeddings are projected into a shared multimodal space using learnable projection matrices, resulting in a 1024-dimensional output representation.  
We employ the music-specific variant of CLAP provided in the official repository at \url{https://github.com/LAION-AI/CLAP}.

\paragraph{LARP \citep{salganik_larp_2024}} addresses the cold-start problem in playlist continuation through a three-stage contrastive learning framework. Built upon the BLIP architecture, LARP consists of two unimodal encoders: HTS-AT~\citep{chen2022htsat} for audio encoding and BERT for text processing, where [CLS] token embeddings are used to represent text. The original 768-dimensional embeddings from both encoders are projected into a unified 256-dimensional embedding space. The framework then performs within-track contrastive learning, track–track contrastive learning, and track–playlist contrastive learning, optimizing representations from both semantic and intra-playlist music relevance perspectives.  
We use the official implementation available at \url{https://github.com/Rsalganik1123/LARP}.

\paragraph{ImageBind \citep{girdhar_imagebine_2023}} unifies six modalities (including image, audio, and text) within a single embedding space through multimodal contrastive learning. Although not music-specific, its general-purpose audio–text alignment capability provides a strong baseline for cross-domain retrieval. ImageBind employs Transformer-based architectures for all modality encoders. For audio input, it converts 2-second audio samples at 16\,kHz into spectrograms using 128 Mel-frequency bins. Treating these spectrograms as 2D signals analogous to images, the model processes them using a Vision Transformer (ViT) with a patch size of 16 and a stride of 10. For text input, ImageBind utilizes pretrained text encoders (302M parameters) from OpenCLIP~\citep{cherti2023openclip}. After projection, all modalities are encoded into a shared 768-dimensional embedding space.  
We extract audio embeddings using the ViT-B/16 variant from the official implementation at \url{https://github.com/facebookresearch/imagebind}.

\paragraph{CLaMP3 \citep{wu_clamp3_2025}} establishes a unified multilingual music–text embedding space by aligning sheet music, audio recordings, and textual descriptions across 12 languages. For audio processing, CLaMP3 adopts pretrained acoustic representations from MERT-v1-95M~\citep{li_mert_2024}. Each 5-second audio clip is represented by an embedding obtained by averaging features across all MERT layers and time steps. For textual content, the model employs XLM-R-base~\citep{conneau2020xlmr}, a multilingual Transformer with 12 layers and 768-dimensional hidden states. The framework uses contrastive learning to align multimodal representations and incorporates additional components such as a retrieval-augmented training mechanism to enhance cross-modal associations.  
We use the checkpoints and architecture from the original authors’ implementation at \url{https://sanderwood.github.io/clamp3}, specifically the SaaS variant optimized for audio.

\subsection{Music-to-Text Generation Models}
\label{app_cross_modal_gen}

\paragraph{MU-LLaMA \citep{liu_music_2024}} is a music-specific adaptation of the LLaMA-2-7B architecture that integrates acoustic features extracted by MERT~\citep{li_mert_2024} through LLaMA-Adapter tuning~\citep{zhang2024llamaadapterefficientfinetuninglanguage}. We use the official implementation provided at \url{https://github.com/shansongliu/MU-LLaMA}, following the same hyperparameter settings as reported by the authors. Specifically, the input audio is split into 60-second segments at a sampling rate of 16\,kHz. The temperature for LLaMA-2-7B is set to 0.6, \textit{top\_p} is set to 0.8, and the maximum sequence length is 1024 tokens.

\paragraph{LP-MusicCaps \citep{doh_lpmusiccaps_2023}} employs a BART-based encoder–decoder architecture~\citep{lewis_bart_2020} with a hidden width of 768 and six Transformer blocks for both the encoder and the decoder. The encoder processes log-Mel spectrograms using convolutional layers similar to those in Whisper~\citep{radford2022robustspeechrecognitionlargescale}. We use the official implementation available at \url{https://github.com/seungheondoh/lp-music-caps} along with the authors’ pretrained checkpoint. For inference, test audio is split into 10-second segments at 16\,kHz, and the longest generated caption among all segments is selected as the final output. In addition, \textit{num\_beams} is set to 5 and the maximum sequence length is 128 tokens.

\paragraph{FUTGA~\citep{wu_futga_2024}} enables time-located music captioning by automatically detecting functional segment boundaries. Built upon SALMONN-7B~\citep{tang_salmonn_2024} with LoRA-based instruction tuning, the model integrates a music feature extractor to support full-length music captioning. For our evaluation, we use the checkpoints and architecture released by the original authors at \url{https://huggingface.co/JoshuaW1997/FUTGA}. In the implementation, Vicuna-7B~\citep{vicuna2023} serves as the language backbone. The repetition penalty is set to 1.5, \textit{num\_beams} is set to 5, \textit{top\_p} is set to 0.95, \textit{top\_k} is set to 50, and each audio input is processed as a 240-second signal sampled at 16\,kHz.

\subsection{Text-to-Music Generation Models}
\label{app_cross_modal_gen-text}

\paragraph{MusicLM \citep{agostinelli_musiclm_2023}} is a generative model that produces high-quality music from text prompts using a hierarchical sequence-to-sequence approach. It leverages audio embeddings from a self-supervised model and autoregressively generates both semantic and acoustic tokens. Unfortunately, this model does not provide publicly available architectures or checkpoints. We therefore use a crowd-sourced implementation available at \url{https://github.com/zhvng/open-musiclm}. Notably, this implementation deviates from the original formulation by using an open-source version of CLAP~\citep{wu_largescale_2023} instead of MuLan~\citep{huang2022mulanjointembeddingmusic}, and EnCodec~\citep{defossez_high_2022} instead of SoundStream~\citep{zeghidour_soundstream_2021}. The purpose of including this implementation is to showcase the performance of a broad range of publicly available models.

\paragraph{Stable Audio \citep{evans_fast_2024}} is a diffusion-based music generation model that synthesizes audio from text prompts and optional melody input using a latent audio representation. The model is built around a latent diffusion framework composed of a variational autoencoder (VAE), a textual conditioning signal, and a diffusion model. The VAE consists of a Descript Audio Codec~\citep{kumar_high_2023} encoder–decoder pair. Textual conditioning is provided by a pretrained CLAP model~\citep{wu_largescale_2023}, specifically the HT-SAT~\citep{chen2022htsat} audio encoder and a RoBERTa-based~\citep{liu2019robertarobustlyoptimizedbert} text encoder. The diffusion component is implemented as a U-Net~\citep{schneider_mousai_2024} with four levels of downsampling encoder blocks and upsampling decoder blocks, connected via skip connections. For our evaluation, we use the checkpoints and architecture released by the original authors at \url{https://github.com/Stability-AI/stable-audio-tools}.

\paragraph{MusicGen \citep{copet_simple_2023}} is a Transformer-based model that generates music from text descriptions. In our experiments, we use the 300M-parameter variant. The model employs a five-layer EnCodec architecture for 32\,kHz monophonic audio with a stride of 640, resulting in a frame rate of 50\,Hz, an initial hidden size of 64, and a final embedding size of 640. The embeddings are quantized using residual vector quantization (RVQ) with four quantizers, each having a codebook size of 2048. During inference, the model uses top-$k$ sampling with $k=250$ and a temperature of 1.0. For evaluation, we use the checkpoints and architecture provided by the original authors at \url{https://github.com/facebookresearch/audiocraft}.

\paragraph{AudioLDM2 \citep{liu_audioldm2_2024}} is a diffusion-based text-to-audio generation model trained on large-scale data and designed to handle diverse audio types, including music and sound effects. It extends prior AudioLDM models by incorporating higher-quality representations and more efficient training strategies. For our evaluation, we use the checkpoints and architecture released by the original authors at \url{https://github.com/haoheliu/AudioLDM2}. Specifically, we adopt the version with a two-layer latent diffusion model. For audio encoding, AudioLDM2 employs an AudioMAE encoder with a patch size of $16\times16$ and no overlap, producing a 768-dimensional feature sequence of length 512 for every 10 seconds of Mel spectrogram input. For text encoding, the model uses a GPT-2 architecture with 12 Transformer layers and a hidden dimension of 768.

\paragraph{Mustango \citep{melechovsky_mustango_2024}} is a multi-stage latent diffusion model for text-to-music generation that emphasizes both musical coherence and audio quality. It introduces a time-aware Transformer to model long audio sequences and supports multi-track generation. For our evaluation, we use the checkpoints and architecture released by the original authors at \url{https://github.com/AMAAI-Lab/mustango}. During inference, the model employs two Transformer-based text-to-music feature generators that predict beat and chord information. Beat prediction is performed using a DeBERTa-Large model~\citep{he2023debertav3improvingdebertausing}, which predicts both meter and inter-beat interval durations, while chord prediction is handled by a FLAN-T5-Large model~\citep{chung_scaling_2022}.

\paragraph{Mureka} is a proprietary music generation model accessible via \url{https://www.mureka.ai}. We implement a custom pipeline for interacting with the Mureka API. This pipeline will be released on our GitHub repository once the current API-related issues are resolved.

\subsection{Dataset Splits}
\label{ap:dataset_splits}

For all evaluations conducted on MusicCaps and Song Describer, we evaluate models on the entirety of the data that is currently publicly available. This choice is motivated by the fact that neither dataset provides officially released or standardized train–test splits that can be consistently used across models. For instance, although the original MusicCaps paper references the existence of a test set, the publicly available version of the dataset released on Kaggle contains only a training split. As a result, many prior works evaluating on MusicCaps construct synthetic test sets by defining their own train–test splits over the available data~\citep{melechovsky_mustango_2024,wu_futga_2024,doh_lpmusiccaps_2023}. Without access to a shared held-out test set or leaderboard, it is therefore not possible to reliably assess or compare performance across studies, nor to fully account for potential overfitting.

A similar situation applies to Song Describer: the dataset, as released, does not include an explicit demarcation between training and evaluation splits, leading each study to adopt its own splitting strategy. Since our work does not involve any fine-tuning on these datasets, we opted to evaluate models on the full publicly available sets in order to report their overall performance. In contrast, MusicSem includes a clearly defined and human-validated test set. For all evaluations on MusicSem, we use only this held-out portion of the data for testing, while releasing the remaining entries as the public training set.

\subsection{Computational Resources and Runtime Analysis}

\paragraph{Computational Resources.}
For generative tasks, all experiments were conducted on systems equipped with NVIDIA L40 GPUs, each providing 48\,GB of VRAM, and using CUDA~12.6. Each experiment was executed on a single GPU instance. For retrieval tasks, all experiments were conducted on systems equipped with NVIDIA A40 GPUs with 46\,GB of VRAM per card, using CUDA~12.4. Similarly, each retrieval experiment was executed on a single GPU instance.

\paragraph{Runtime Analysis.}
For text-to-music generation, we analyze the relationship between inference time and the duration of the generated audio in Table~\ref{tab:latency_t2m}. Because generation duration varies substantially across models, and producing long, coherent musical segments remains a central challenge in text-to-music generation~\citep{copet_simple_2023}, we evaluate each model using the duration settings specified in its original formulation and codebase.

We report a trade-off metric computed as the ratio of inference time to generation duration. From the results in Table~\ref{tab:latency_t2m}, we observe that among publicly available models, Stable Audio achieves the lowest inference latency. In addition, the proprietary model Mureka is able to generate longer stretches of coherent audio than all publicly available models, highlighting a clear performance gap between open-source and commercial generation systems.

For music-to-text generation, we evaluate the relationship between inference time and the length of the generated textual annotation in Table~\ref{tab:latency_m2t}. The results indicate that LP-MusicCaps achieves the highest trade-off, meaning that one second of inference time yields the largest number of generated characters.

Finally, for text-to-music retrieval, we evaluate the inference latency of cross-modal retrieval models in Table~\ref{tab:latency_retrieval}. As shown by the results, inference latency varies only marginally across models, although ImageBind~\citep{girdhar_imagebine_2023} is slightly faster than the other approaches.

\begin{table}[t]
\centering
\caption{Inference time of text-to-music generation models on MusicSem. \text{Trade-off} is defined as inference time divided by generation size.}
\label{tab:latency_t2m}
\resizebox{\linewidth}{!}{
\begin{tabular}{c|ccc}
\toprule 
\textbf{Model} & \textbf{Inference Time (in seconds)} & \textbf{Generation Size (in seconds)} & \textbf{Tradeoff} ↓ \\
\midrule  
MusicLM       & 102  & 5   & 20.40 \\
AudioLDM2     & 13   & 10  & 1.30 \\
Mustango      & 50   & 10  & 5.00 \\
MusicGen      & 40   & 20  & 2.00 \\
Stable Audio  & 18   & 45  & 0.40 \\
Mureka        & 120  & 150 & 0.80 \\
\bottomrule
\end{tabular}}
\end{table}

\begin{table}[t]
\centering
\caption{Inference time of music-to-text generation models on MusicSem. \text{Trade-off} is defined as inference time divided by generation size.}
\label{tab:latency_m2t}
\resizebox{\linewidth}{!}{
\begin{tabular}{c|ccc}
\toprule 
\textbf{Model} & \textbf{Inference Time (in seconds)} & \textbf{Generation Size (in characters)} & \textbf{Tradeoff} ↓ \\
\midrule  
LP-MusicCaps   &   8& 2000   & 0.004 \\
MU-LLaMA    & 4 &  70 &0.057   \\
FUTGA      & 15   & 1138 & 0.013 \\
\bottomrule
\end{tabular}
}
\end{table}
\begin{table}[t]
\centering
\caption{Inference time of cross-modal retrieval models on MusicSem.}

\label{tab:latency_retrieval}
\resizebox{0.45\linewidth}{!}{
\begin{tabular}{c|c}
\toprule 
\textbf{Model} & \textbf{Inference Time (in seconds)} \\
\midrule  
LARP   &   0.26  \\
CLAP    & 0.23  \\
ImageBind      & 0.21  \\
CLAMP3      & 0.28  \\
\bottomrule
\end{tabular}
}
\end{table}

\subsection{Evaluation Metrics}
\label{app:metrics}

\paragraph{Interpreting Music-to-Text Metrics.} In this section, we provide a brief overview of the evaluation metrics used for assessing music-to-text generation models. Following canonical works in music-to-text generation~\citep{liu_music_2024,doh_lpmusiccaps_2023}, we first consider three $n$-gram-based metrics originally developed for machine translation: BLEU~\citep{papineni_bleu_2002}, ROUGE~\citep{lin_rouge_2004}, and METEOR~\citep{banerjee_meteor_2005}. 

BLEU (B) measures precision by computing the overlap of $n$-grams (typically unigrams, bigrams, and trigrams; i.e., B$_1$, B$_2$, and B$_3$) between the reference annotation and the generated music caption. In contrast, ROUGE (R) emphasizes recall by measuring the overlap of $n$-grams between the generated caption and the reference annotation. METEOR (M) is designed to better align with human judgment by extending exact $n$-gram matching to include synonymy and paraphrase-based matches, thereby addressing some limitations of BLEU and ROUGE.

We also include CIDEr~\citep{vedantam_cider_2015}, a metric originally proposed for image captioning, which measures the similarity between a generated caption and a set of reference annotations using a weighted $n$-gram scheme that emphasizes consensus. Finally, we report BERTScore~\citep{bert}, which compares contextualized embeddings of generated and reference captions using a pretrained BERT model, thereby capturing semantic similarity beyond surface-level lexical overlap.

The purpose of employing this diverse set of metrics is to capture increasing levels of abstraction in evaluating the alignment between original annotations and generated captions. As observed in our experiments, BERTScore tends to be the most stable across datasets, whereas $n$-gram-based metrics exhibit higher variability across both datasets and models.

\paragraph{CLAP Score.} The Contrastive Language–Audio Pretraining~\citep{wu_largescale_2023} score (CLAP score) is a simple yet effective, reference-free metric that quantifies how well an audio signal aligns with a textual description. This metric is commonly used in text-to-music generation to evaluate how accurately a generative model expresses the information provided in a textual prompt.
Formally, given a set of paired textual inputs and generated audio outputs $(T, \tilde{A})$, where the audio $\tilde{A} = \mathcal{M}(T)$ is generated by conditioning a music generation model $\mathcal{M}$ on the textual input $T$ (e.g., MusicGen~\citep{copet_simple_2023}), embeddings for each modality are computed using the CLAP model as follows:
\begin{equation}
    Z_{\tilde{A}} = \text{CLAP}_{\text{audio}}(\tilde{A}), \qquad
    Z_T = \text{CLAP}_{\text{text}}(T),
\end{equation}
where $Z_{\tilde{A}}$ and $Z_T$ denote the audio and text embeddings produced by the CLAP audio and text encoders, respectively.
Given these embeddings, the CLAP score is computed as the average cosine similarity between corresponding audio and text representations in the shared embedding space. Using matrix-style indexing, where $Z_{\tilde{A}}[i]$ denotes the $i$-th audio embedding and $Z_T[i]$ the corresponding text embedding, the CLAP score is defined as:
\begin{equation}
    CS(T, \tilde{A}) = \frac{1}{n} \sum_{i=1}^{n}
    \frac{\langle Z_{\tilde{A}}[i], Z_T[i] \rangle}
    {\|Z_{\tilde{A}}[i]\| \cdot \|Z_T[i]\|}.
\end{equation}
Intuitively, higher CLAP scores indicate stronger alignment between the audio and textual representations in the joint embedding space.

\paragraph{Vendi Score.} It is not immediately obvious that the Vendi score \citep{friedman_vendi_2023}, which was originally proposed for images, is directly applicable to audio spectrograms, i.e., image-like representations of audio signals in the frequency domain. To assess whether the Vendi Score is sensitive to meaningful variations in music, we conduct an ablation study. We consider 15 seed tracks. For each seed track, we select three \textit{positive} and three \textit{negative} examples. Positive examples correspond to cover songs, in which different musicians perform the same musical piece as the original seed track. In contrast, negative examples consist of songs from entirely different genres and artists. We hypothesize that, if the Vendi score can effectively measure diversity in collections of audio, it should clearly distinguish between positive and negative examples when evaluated relative to a  seed~track.

\begin{figure}[t]
    \centering
    \includegraphics[width=0.9\linewidth]{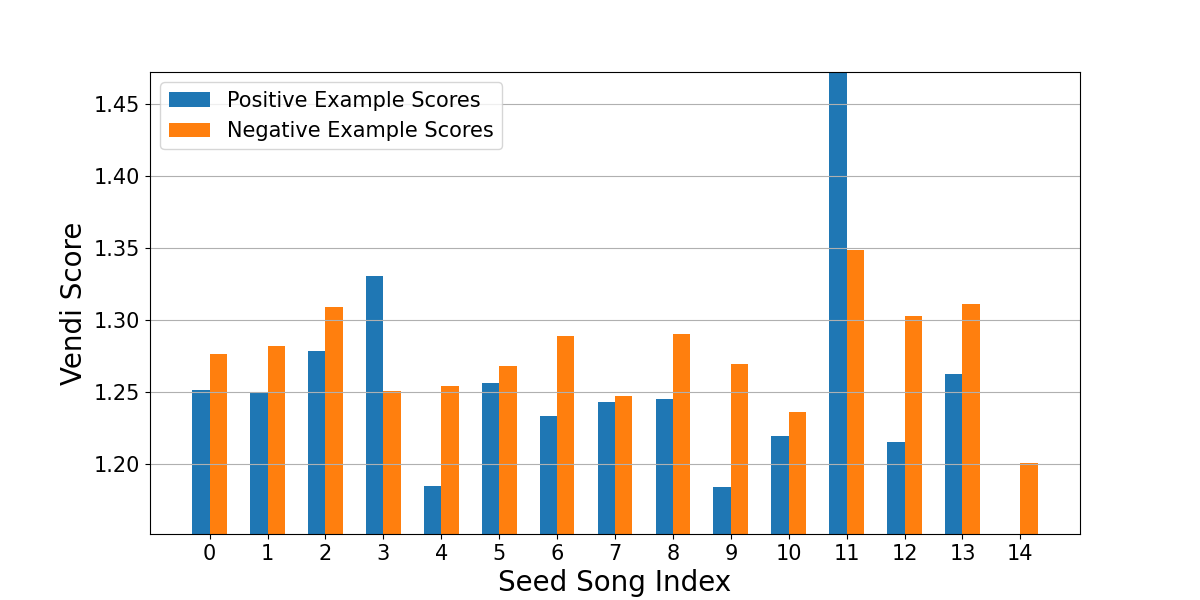}
    \caption{Comparison between Vendi scores for cover songs and random collection.}
    \label{fig:cover_songs}
\end{figure}

As shown in Figure~\ref{fig:cover_songs}, the Vendi Score is indeed able to differentiate between these groups. For nearly all seed tracks (shown along the x-axis), the score is consistently higher for negative examples (orange) than for cover songs (blue), indicating greater diversity relative to the seed track. This result suggests that the Vendi Score captures meaningful differences in musical content and can be reasonably applied as a diversity metric for audio. The list of songs used in this ablation study is available at \url{https://tinyurl.com/2ff3d4f6}.

\end{document}